\documentclass[12pt, draftclsnofoot, onecolumn]{IEEEtran}

\IEEEoverridecommandlockouts

\usepackage{graphicx,cite,acronym,amsmath,amssymb,amsfonts,color,subfigure,floatflt,stfloats,bm,textgreek}
\usepackage{epsfig,epstopdf,psfrag}

\usepackage[ruled,vlined]{algorithm2e}
\usepackage{setspace}

\usepackage{soul}
\usepackage{comment}

\usepackage{comment}
\usepackage{soul}

\SetKwInput{KwInput}{Input}                
\SetKwInput{KwOutput}{Output}  

\usepackage[table]{xcolor}
\usepackage{bm}

\usepackage[cmintegrals]{newtxmath}
\usepackage{graphics,hhline,array,cite,bbm}

\usepackage{latexsym}
\usepackage{theorem}
\usepackage{times}
\usepackage{makeidx}

\DeclareGraphicsExtensions{.png,.jpg,.eps}

\acrodef{IIoT}{industrial Internet-of-things}

\acrodef{IoT}{Internet-of-things}

\acrodef{AP}{access point}

\acrodef{ABIM}{active backscattering intelligent metasurface}

\acrodef{SVD}{singular-value decomposition}

\acrodef{EVD}{eigen-value decomposition}

\acrodef{PSWF}{prolate spheroidal wave function}

\acrodef{PER}{packet error rate}

\acrodef{CR}{channel response}

\acrodef{BS}{base station}

\acrodef{MS}{mobile station}

\acrodef{UE}{user equipment}

\acrodef{MIMO}{multiple-input multiple-output}

\acrodef{MU-MIMO}{multi-user MIMO}

\acrodef{mMTC}{massive Machine Type Communication}

\acrodef{RIS}{reconfigurable intelligent surface}

\acrodef{IRS}{intelligent reconfigurable surface}

\acrodef{LIS}{large intelligent surface}

\acrodef{MIS}{medium intelligent surface}

\acrodef{SIS}{small intelligent surface}

\acrodef{DoF}{degrees-of-freedom}

\acrodef{AF}{amplify \& forward}

\acrodef{DF}{detect \& forward}

\acrodef{JF}{just forward}

\acrodef{CSI}{channel state information}

\acrodef{RV}{random variable}

\acrodef{i.i.d.}{independent, identically distributed}

\acrodef{PSD}{power spectral density}

\acrodef{PDF}{probability distribution function}

\acrodef{CDF}{cumulative distribution function}

\acrodef{ch.f.}{characteristic function}

\acrodef{AWGN}{additive white Gaussian noise}

\acrodef{RSSI}{received signal strength indicator}

\acrodef{SNR}{signal-to-noise ratio}

\acrodef{SINR}{signal-to-interference-noise ratio}

\acrodef{LRT}{likelihood ratio test}

\acrodef{GLRT}{generalized likelihood ratio test}

\acrodef{GML}{generalized maximum likelihood}

\acrodef{LOS}{line-of-sight}

\acrodef{NLOS}{non-line-of-sight}

\acrodef{GDOP}{geometric dilution of precision}

\acrodef{GPS}{Global Positioning System}

\acrodef{FIM}{Fisher information matrix}

\acrodef{PEB}{position error bound}

\acrodef{WSN}{Wireless Sensor Network}

\acrodef{MAC}{medium access control}

\acrodef{RSS}{received signal strength}

\acrodef{RTT}{round-trip time}

\acrodef{MIMO}{multiple-input multiple-output}

\acrodef{MF}{matched filter}

\acrodef{ED}{energy detector}

\acrodef{ML}{maximum likelihood}

\acrodef{NL}{nonlinear}

\acrodef{MSE}{mean square error}

\acrodef{RMSE}{root mean square error}

\acrodef{ppm}{part-per-million}

\acrodef{PRP}{pulse repetition period}

\acrodef{ACK}{acknowledge}

\acrodef{UWB}{ultrawide bandwidth}

\acrodef{TNR}{threshold-to-noise ratio}

\acrodef{NLOS}{non line-of-sight}

\acrodef{LOS}{line-of-sight}

\acrodef{LS}{least squares}

\acrodef{IR-UWB}{impulse radio UWB}

\acrodef{FCC}{Federal Communications Commission}

\acrodef{TH}{time-hopping}

\acrodef{PPM}{pulse position modulation}

\acrodef{PAM}{pulse amplitude modulation}

\acrodef{MUI}{multi-user interference}

\acrodef{PDP}{power delay profile}

\acrodef{PPP}{Poisson point process}

\acrodef{DS}{delay spread}

\acrodef{CED}{channel excess delay}

\acrodef{BPZF}{band-pass zonal filter}

\acrodef{SIR}{signal-to-interference ratio}

\acrodef{RFID}{radio frequency identification}

\acrodef{WPAN}{wireless personal area networks}

\acrodef{WWLB}{Weiss-Weinstein lower bound}

\acrodef{DP}{direct path}

\acrodef{MF}{matched filter}

\acrodef{MMSE}{minimum-mean-square-error}

\acrodef{SBS}{serial backward search}

\acrodef{NBI}{narrowband interference}

\acrodef{WBI}{wideband interference}

\acrodef{INR}{interference-to-noise ratio}

\acrodef{CIR}{channel impulse response}

\acrodef{ISI}{inter-symbol interference}

\acrodef{CPR}{channel pulse response}

\acrodef{LRT}{likelihood ratio test}

\acrodef{RADAR}{RADAR}

\acrodef{MUR}{Multistatic RADAR}

\acrodef{MUI}{multi-user interference}

\acrodef{EM}{electromagnetic}

\acrodef{CW}{continuous wave}

\acrodef{RF}{radiofrequency}

\acrodef{FCC}{Federal Communications Commission}

\acrodef{EIRP}{effective radiated isotropic power}

\acrodef{RCS}{radar cross-section}

\acrodef{BAV}{balanced antipodal Vivaldi}

\acrodef{PRake}{partial Rake}

\acrodef{RTLS}{real time locating system}

\acrodef{CRB}{Cram\'{e}r-Rao bound}

\acrodef{ZZB}{Ziv-Zakai bound}

\acrodef{TOA}{time-of-arrival}

\acrodef{TOF}{time-of-flight}

\acrodef{WSN}{wireless sensor network}

\acrodef{MAC}{medium access control}

\acrodef{RSS}{received signal strength}

\acrodef{TDOA}{time difference-of-arrival}

\acrodef{RF}{radiofrequency}

\acrodef{RTT}{round-trip time}

\acrodef{AOA}{angle-of-arrival}

\acrodef{MF}{matched filter}

\acrodef{ED}{energy detector}

\acrodef{ML}{maximum likelihood}

\acrodef{MUR}{Multistatic radar}

\acrodef{HDSA}{high-definition situation-aware}

\acrodef{RRC}{root raised cosine}

\acrodef{OFDM}{orthogonal frequency division multiplexing}

\acrodef{IF}{intermediate frequency}

\acrodef{PHY}{physical layer}

\acrodef{S-V}{Saleh-Valenzuela}

\acrodef{UHF}{ultra-high frequency}

\acrodef{PR}{pseudo-random}

\acrodef{SoC}{System on Chip}

\acrodef{SoP}{System on Package}

\acrodef{SPMF}{Single-Path Matched Filter}

\acrodef{IMF}{Ideal Matched Filter}

\acrodef{SCR}{signal-to-clutter ratio}

\acrodef{BEP}{bit error probability}

\acrodef{BER}{bit error rate}

\acrodef{WSR}{wireless sensor radar}

\acrodef{HPBW}{half power beam width}

\acrodef{LEO}{localization error outage}

\acrodef{WSS}{wide-sense stationary}

\acrodef{TR}{time-reversal}

\acrodef{WSSUS}{WSS with uncorrelated scattering}

\acrodef{GP}{Gaussian process}

\acrodef{IMU}{inertial measurement unit}

\acrodef{TDD}{time-division duplexing}

\acrodef{ULA}{uniform linear array}

\acrodef{URLLC}{ultra-reliable low-latency communications}

\acrodef{SCM}{self-conjugating metasurface}

\acrodef{CDL}{Clustered Delay Line}

\newcommand{\boldA} {{\bf{A}}}

\newcommand{\bolds} {{\bf{s}}}

\newcommand{\boldb} {{\bf{b}}}

\newcommand{\boldu} {{\bf{u}}}

\newcommand{\boldH} {{\bf{H}}}

\newcommand{\boldC} {{\bf{C}}}
\newcommand{\boldB} {{\bf{B}}}

\newcommand{\boldI} {{\bf{I}}}
\newcommand{\boldr} {{\bf{r}}}

\newcommand{\boldn} {{\bf{n}}}
\newcommand{\boldx} {{\bf{x}}}
\newcommand{\boldy} {{\bf{y}}}

\newcommand{\boldz} {{\bf{z}}}
\newcommand{\boldw} {{\bf{w}}}
\newcommand{\boldv} {{\bf{v}}}

\newcommand{\fc} {f_{\text{c}}}

\newcommand{\rank}[1]{{\rm rank} \left ( #1 \right )}

\newcommand{\boldeta} {{\boldsymbol{\eta}}}

\newcommand{\Ptx} {P_{\text{T}}}

\newcommand{\Pscout} {P_{\text{scout}}}

\newcommand{\Gbs} {G_{\text{BS}}}
\newcommand{\Gscm} {G_{\text{SCM}}}

\newcommand{\Fbs} {F_{\text{BS}}}
\newcommand{\Fscm} {F_{\text{SCM}}}

\newcommand{\oy} {\mathring{\bf{y}}}
\newcommand{\ov} {\mathring{\bf{v}}}

\newcommand{\oA} {\mathring{\bf{A}}}
\newcommand{\orank} {\mathring{r}}
\newcommand{\olambda} {\mathring{\lambda}}

\begin{document}
\title{Grant-free Random Access with Self-conjugating Metasurfaces}

\author{
\IEEEauthorblockN{Davide~Dardari,~\IEEEmembership{Senior~Member,~IEEE},   
 Marina~Lotti,~\IEEEmembership{Student~Member,~IEEE}, Nicol\'o  Decarli,~\IEEEmembership{Member,~IEEE}, Gianni Pasolini,~\IEEEmembership{Member,~IEEE}
}
}

\maketitle

\begin{abstract}
Recently, grant-free random access schemes have received significant attention in the scientific community as a solution for extremely low-latency massive communications in new  \ac{IIoT} and digital twins applications. Unfortunately, the adoption of such schemes in the mmWave and THz bands is challenging because massive antenna arrays are needed to counteract the high path loss and provide massive access with consequent significant signaling overhead for channel estimation and slow beam alignment procedures between the \ac{BS} and \acp{UE}, which are in  contrast to the ultra-low-latency requirement, as well as to the need for low hardware complexity and energy consumption. 
In this paper, we propose the adoption of a \ac{SCM}  at the \ac{UE} side, where the signal sent by the \ac{BS} is reflected after being conjugated and phase-modulated according to the \ac{UE} data.
Then, a novel grant-free random access protocol is presented capable to detect new \acp{UE} and  establish parallel  \ac{MIMO} uplink communications  with almost zero latency and jitter.
This is done in a blind way without the need for RF/ADC chains at the \ac{UE} side as well as without explicit channel estimation and time-consuming beam alignment schemes. 
\end{abstract}

\begin{IEEEkeywords}
Self-conjugating metasurfaces; Grant-free random access; MIMO; Retro-directive backscattering; Intelligent surfaces. 
\end{IEEEkeywords}

\newpage

\section{Introduction}

Future wireless networks are expected to satisfy very challenging requirements in terms of data rate, latency, energy efficiency, and nodes density, in order to enhance the performance of existing \ac{mMTC} applications and enable new ones \cite{PokDinJihCho:20}.
Among them, the creation of digital twin worlds perfectly intertwined with physical objects in  \acf{IIoT} environments, is certainly one of the most challenging  
\cite{VisHarMogPre:20,WanGaoDiRHan:22}. 
Here, ultra-low latency ($<100\,u$s), low jitter ($\approx 1-2\,u$s), ultra-massive access (10 sensors/m$^2$), high energy efficiency, and low complexity, often must  be satisfied simultaneously. 
The exploitation of high frequencies in the mmWave and THz bands is one direction of investigation to gain more bandwidth, which however requires massive antenna arrays with very narrow beams (also at the user side) to compensate for the high path loss   and obtain better spatial filtering to reduce the interference between users 
 that are using the same radio resource. 
 However, massive antenna arrays also introduce new issues due to  signaling overhead (e.g., for channel estimation), high hardware complexity, increased power consumption, and slow beamforming and alignment procedures, which are in contrast with the low-latency requirement and the possibility to embed wireless devices into low-cost, small and low-energy sensors (e.g., exploiting energy harvesting techniques \cite{CosDarAleDecDelFabFanGueMasPizRom:J17}).
In addition, in \acp{mMTC} the generated traffic is often sporadic, random, and characterized by the transmission of short packets. 
Under these conditions, to keep the access latency at extremely low values, any coordination action from the \ac{BS}, such as scheduling, synchronization,  or retransmission, is typically avoided in favor of grant-free access schemes \cite{CheNgYuLarAlDSch:21}. In this context, \ac{CSI} estimation becomes extremely challenging, especially in \ac{MIMO} systems. In fact,  channel estimation typically  relies on the assumption that devices are active for a long period so that long preambles (pilot symbols) can be allocated without a significant loss in efficiency. Vice versa,  in \acp{mMTC} the \ac{CSI}
 has to be estimated on a one-shot basis with an extremely limited space available for pilot symbols allocation.    
 For instance, the initial beam alignment procedure (i.e., the \ac{CSI}  estimation) requires, in general, that both the \ac{BS} and \ac{UE} test all possible pairs of angles before the \ac{UE} could be visible to the \ac{BS} and the communication established \cite{GioMezZor:16}.   
These approaches either involve some computational effort on both sides, which affects hardware complexity and power consumption, or long training data/search time, which increases latency. Clearly, these problems are emphasized when the number of antennas is high. Moreover, at high frequencies, such as sub-THz/THz, technological constraints place several limitations on the flexibility in processing the signal sent/received by each individual antenna element, thus reducing the set of affordable solutions (digital bottleneck). 
Therefore, it is of paramount importance to study solutions enabling ultra-low latency grant-free \ac{MIMO} random access using low-complexity, low-power devices at mmWave and beyond,  characterized by limited or no overhead associated with \ac{CSI} estimation and user detection.  

Recently, intelligent metasurfaces have attracted a lot of interest in the research community as a technology to realize extremely (electrically) large antennas, namely \acp{LIS}, and \acp{RIS} capable to modify the way the \ac{EM} waves propagate in the environment with zero additional latency and low-consumption \cite{QuYiCheNgCha:17,BarHamLonMonRamVelAleBil:21,DiRZapDebAloYueRosTre:20,FaeMinGonCamMarDelMac:19}.
Typically, \acp{RIS} have been proposed to cope with coverage extension in \ac{NLOS} channel conditions \cite{MurSciGarCotPerGes:21,DarDec:J21,DarMas:J21,AbrDarDiR:J21}.
Other recent uses  consider the possibility for the \ac{RIS} to embed opportunistically some information while reflecting a signal from a \ac{BS} towards the intended user(s) \cite{HuZhaDiBiaSon:22,DinAloXinLiXu:22,LiaZhaWanLonZhoYan:22}. This is done through the adoption of backscatter modulation  \cite{DarDErRobSibWin:J10,NiuLi:19}.   
Another interesting direction of research to overcome the digital bottleneck is to move some basic processing functionalities  directly to the intelligent metasurface, that is, at  the \ac{EM} level. This is expected to bring advantages in terms of latency, power consumption, size, and complexity reduction, especially at mmWave and THz, with respect to the digital implementation counterpart \cite{TarEle:21,Sil:14}. 
One question is therefore whether intelligent metasurfaces can  help in finding solutions to the above-mentioned challenges for grant-free \ac{MIMO} random access. 

For the purposes of this paper, we focus on a particular type of \ac{RIS}, usually denoted as \acf{SCM}. \acp{SCM} are specific surfaces capable of reflecting an \ac{EM} wave, typically employed to obtain the retro-directivity effect \cite{MiyIto:02}, whose complex envelope is the conjugate of the envelope of the impinging \ac{EM} wave. 
Retro-directivity has been investigated for a long time in different contexts such as radar, wireless power transfer, collision avoidance systems, microwave imaging or detection, RFID systems, and remote information retrieval from sensors. In \cite{DarLotDecPas:C23}, an \ac{SCM} is used to continuously track a single user in a fast time-variant channel. 
However, to the Authors' best knowledge, its use for efficient wireless communications has been little studied \cite{MiyIto:02} and has never been considered for grant-free random access \ac{MIMO} systems,  as we do in this work.    
Specifically, in this paper, we exploit the properties of \acp{SCM} to design an almost zero-latency grant-free random access \ac{MIMO} scheme with asynchronous \acp{UE}. One important peculiarity of the proposed scheme is that it does not need any overhead for \ac{CSI} estimation and  any ADC and baseband processing at \ac{UE} side, thus allowing the efficient transmission of short packets in \ac{mMTC} applications with low-complexity devices.

\subsection{Related Works}

Massive random access has been a hot topic for many years and several approaches have been proposed. For a general overview, the reader can refer to the survey in \cite{CheNgYuLarAlDSch:21}.  
Grant-free massive random access schemes  have to deal not only with  inter-user interference but also with the problem of detecting the presence of active users and estimating the channel \cite{LiuYu:18,LiuLarYuPopSteDeC:18}. 
Various solutions have been proposed. For instance, coded random access protocols allow an increased level of reliability thanks to the joint exploitation of interference cancellation and coding techniques on top of classical ALOHA-based random access protocols \cite{PaoSteLivPop:15}. As an example, explicit preamble transmission is avoided in the coded random access scheme investigated in \cite{YanFanLiDinHao:22} by assigning to each user a unique channel access pattern  which is used to embed user identities and facilitate the detection. In any case, coded random access relies on the transmission of some replicas of the same packet in order to perform interference cancellation that might deny the achievement of sub-ms latency. 
In \cite{MohDobWin:22}, a single-antenna massive access scheme is proposed where the device ID is embedded in the sequence used to spread the signal, thus eliminating the explicit transmission of the ID,  and the channel estimation is avoided thanks to  a non-coherent multi-user detector aided by an unsupervised machine learning algorithm. The main drawbacks of the scheme are the increased bandwidth due to spreading and the fact that it is tailored to  single-antenna configurations.      

Massive \ac{MIMO} has demonstrated the possibility to reduce the effect of inter-user interference to any level provided that the number of antennas at the \ac{BS} grows indefinitely under typical channel characteristics \cite{WuLiu:17}. This allows  communication with  a large number of users simultaneously sharing the same radio resource.   
In \cite{LiuYu:18}, an  asymptotic regime analysis of user activity detection and channel estimation is carried out as the number of antennas at the \ac{BS} becomes massive. Specifically, the authors quantify the performance of user activity detection and channel estimation when randomly generated non-orthogonal pilot sequences are assigned to each device. They show that  the user activity detection error can be arbitrarily small when increasing the number of antennas, whereas the channel estimation error and the overhead for pilots remain the main bottleneck. 
The rich dimensionality of the sparse angular domain \ac{MIMO} channel is exploited in \cite{XieWuAnGaoZhaXinWonXia:22}
to design an uncoupled slotted data transmission scheme for  unsourced random access. The similarity of the angular transmission pattern implied
in the slot-wise reconstructed channels allows the design of  a clustering-based decoder combining message sequences across slots.
%
The work in  \cite{BanXuCarPop:18} combines instantaneous \ac{CSI}
with long-term statistical {CSI} in a massive \ac{MIMO} context  assuming that channel singular vectors are  known and constant over a long term. The beamforming design is shown to improve the mean \ac{SNR} and reduce its standard deviation while keeping the latency below $0.5\,$ms,  and hence suitable for \ac{URLLC} applications.
Finally, in \cite{XiuGaoLiaMeiZheTanDiRHan:22},   a joint active user detection and channel estimation scheme  in support of massive access is proposed with the adoption of a multi-panel massive \ac{MIMO} in a cellular \ac{IIoT} scenario working at mmWave and THz. The scheme takes advantage of the sparsity of active users to successfully exploit compressed sensing techniques. 

Despite the rich literature on the subject, to the authors' knowledge, no solutions have been proposed  ensuring grant-free random access with ultra-low latency (e.g., $<100\,\mu$s) and jitter, no overhead for \ac{CSI} estimation, in a \ac{MIMO} \ac{mMTC} scenario using ultra-low complexity devices. 


 
  

\subsection{Our Contribution}

In this paper, we introduce a  \ac{MIMO} scheme which overcomes the previous limitations and allows zero-latency grant-free access to asynchronous \ac{MIMO} devices equipped with \acp{SCM}.
The main idea is to have different parallel tasks running at the \ac{BS} that generate proper sounding signals and process  the signal received by the \ac{BS} antenna array. 
In order to detect new users, a dedicated \emph{Scouting Task} 
is always running on a channel subspace orthogonal to those  used by the already active links. 
In particular, the Scouting Task jointly exploits retro-directivity and the backscatter communication of  a modulating  \ac{SCM} at the \ac{UE} side, where the signal sent by the \ac{BS} is reflected (backscattered) after being conjugated and phase-modulated according to the user's data directly at \ac{EM} level (\emph{retro-directive backscattering}). It will be shown that operating in this way, the \ac{BS} and the \ac{UE} can establish a \ac{MIMO} uplink communication  without any computational effort at the \ac{UE} side nor any signaling.  More precisely, thanks to the methodology introduced in this paper, the \ac{BS} is able to derive the optimal beamforming vector (precoding vector), which corresponds to the top eigenvector of the \ac{BS}-\ac{UE} round-trip channel, by  iterating with the \ac{SCM} and making use of a modified \emph{Power Method} for the eigenvectors estimation \cite{GolVor20}. 
This is done blindly in an almost optimal manner, without the need for either the \ac{CSI} estimation or time-consuming beam alignment schemes.
Every time a new user is detected, the \ac{BS} activates a dedicated task, named \emph{Communication Task}, which makes use of the beamforming vector estimated by the Scouting Task addressing a channel subspace orthogonal to those already under exploitation by the Communication Tasks serving the other currently active users.
The result is a scheme where complexity grows linearly with the number of \acp{UE} and is capable of managing asynchronous random transmissions of packets of any size without the explicit  estimation 
of the multi-user \ac{MIMO} channel, thus ensuring an almost zero  latency  and no jitter independently of the number of \acp{UE}. 

We further characterize analytically the convergence behavior of the Scouting Task taking into account the noise generated by the \ac{SCM} and the \ac{BS}. 
Numerical results put in evidence that almost optimal beamforming can be achieved within the transmission of a few symbols at the \ac{BS} and \ac{UE} sides, without any processing at the \ac{UE}.  The increased path loss due to the backscatter nature of the communication link can be easily compensated by increasing the number of antennas at the \ac{UE} without additional processing. 
The adoption of the \ac{SCM} at the \ac{UE} allows an extremely low-complexity and low-power design of the \ac{UE} because no ADC chains are needed. 
It is worth remarking that in this case, the metasurface is part of the \ac{UE}, not an additional element deployed in the environment like in most of the \ac{RIS}-based scenarios investigated in the literature. 

The rest of the paper is structured as follows: In Sec. \ref{Sec:SCM}, a brief overview of  \acp{SCM} is given and the strategy to exploit them to transmit data is proposed. The illustration of the working principle of \ac{SCM}-based communication is given in Sec. \ref{Sec:Working}. The proposed algorithms for random access are described  in Sec.  \ref{sec:RandomAccess} and the convergence analysis is presented in Sec. \ref{sec:Convergence}. Numerical results are illustrated in Sec. \ref{Sec:NumericalResults}, whereas the conclusions are drawn in Sec. \ref{Sec:Conclusions}.

\subsection{Notation and Definitions}
Boldface lower-case letters are vectors (e.g., $\boldx$), whereas boldface capital letters are matrices (e.g., $\boldH$), where $\boldI_N$ is the identity matrix of size $N$, 
and $\|\boldx \|$ represents the Euclidean norm of vector $\boldx$. $\boldH^T$ and $\boldH^{\dag}$ indicate, respectively, the transpose and the conjugate transpose operators applied to matrix $\boldH$,  whereas $\| \boldH \|_{\text{F}}$ denotes the Frobenius norm of $\boldH$.  
The notation $x \sim {\mathcal{CN}}\left (m, \sigma^2 \right )$ indicates a complex circular symmetric Gaussian \ac{RV} with mean $m$ and variance $\sigma^2$, whereas $\boldx \sim {\mathcal{CN}}\left (\mathbf{m}, \boldC \right )$ denotes a complex Gaussian random vector with mean $\mathbf{m}$ and covariance matrix $\boldC$. 

\begin{figure}[t!]
	\centering \includegraphics[trim= {0 0 0 0}, clip, width=0.6\linewidth]{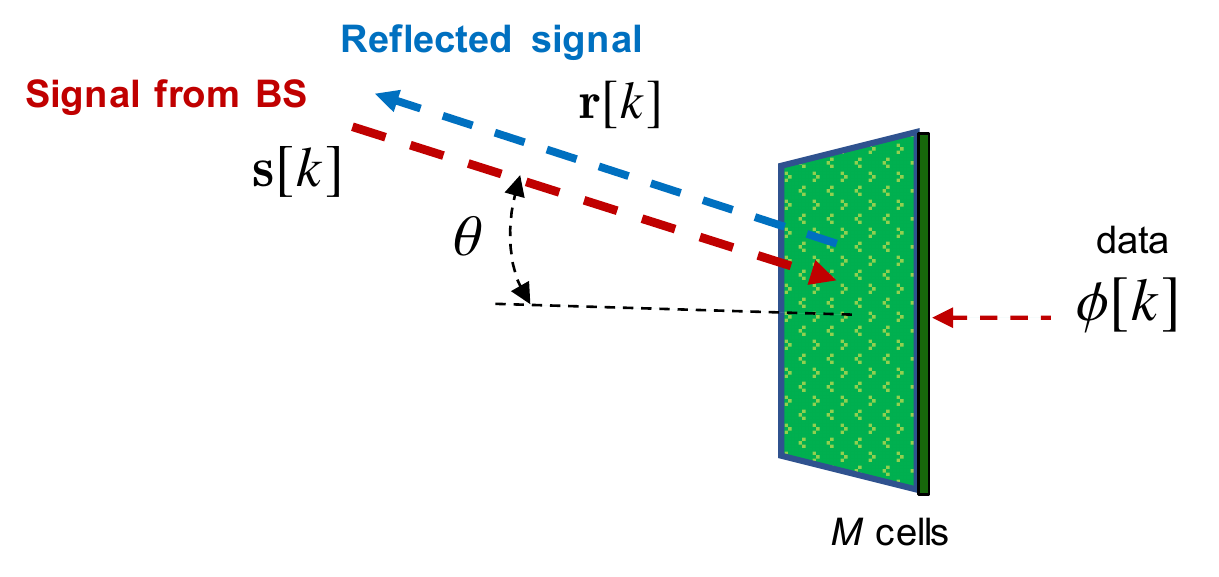}
		\caption{Modulating \acf{SCM} at the \ac{UE} side.}
		\label{Fig:SCM}
\end{figure}

\section{Modulating Self-Conjugating Metasurfaces}
\label{Sec:SCM}

With reference to Fig. \ref{Fig:SCM}, consider a metasurface  (or an antenna array), namely \ac{SCM}, composed of $M$ elements (cells) and a narrowband impinging signal $\bolds[k]=[s_1[k], s_2[k], \ldots, s_M[k]  ]^T$ of bandwidth $W$, where $s_m[k]$ represents the equivalent low-pass discrete-time version of the signal at the input of the $m$th cell at discrete time $k$.
A \ac{SCM} is characterized by the property such that the backscattered equivalent low-pass signal is the conjugate of the input signal  \cite{MiyIto:02}, that is,
\begin{equation}  \label{eq:br1}
\boldr[k]=g\, \bolds^*[k] + g\, \boldeta^*[k]
\end{equation}  
where  $\boldeta[k] \in \mathbb{C}^{M \times 1}$ is the \ac{AWGN}, with $\boldeta[k] \sim {\mathcal{CN}}\left (\mathbf{0}, \sigma_{\eta}^2 \boldI_M \right )$, and $g$ is the gain of the cell ($g<1$ if it is passive). Note that $\sigma_{\eta}^2=\kappa T_0 \Fscm  \, W$, being $\kappa$ the Boltzmann constant, $T_0=290\,$K, and $\Fscm$ the cell's noise figure \cite{BouMagMes:12,KisHum:12,LonGrbHra:19,ZhaDaiCheLiuYanSchPoo:21}.

An interesting property of an \ac{SCM} is \emph{retro-directivity} \cite{MiyIto:02}. In fact, when it is illuminated by a plane wave with incident angle $\theta$, ideally the signal is backscattered  towards the same angle. 
For example, if an \ac{ULA}-like  \ac{SCM} with inter-element spacing $\Delta$ is considered, an impinging plane wave with incident angle $\theta$ can be generically expressed as $s_m[k]=A\, e^{\jmath \frac{2\pi}{\lambda} m\, \Delta \, \sin (\theta)}$, for $m=1,2, \ldots, M$, where $\lambda$ is the wavelength and $A$ is a generic complex constant.
According to \eqref{eq:br1} and neglecting the noise term, the backscattered signal is $r_m[k]=g \, s_m^*[k]=g \, A^* \, e^{\jmath \frac{2\pi}{\lambda} m\,  \Delta \, \sin (-\theta)}$, for $m=1,2, \ldots, M$, which corresponds to a plane wave reflected back toward the same direction $\theta$.    

The most common approach to realize the self-conjugating property is through heterodyne mixing of the incoming wave, centered at frequency $f_0$, with a locally generated sinusoid oscillating at $2f_0$. One of the earliest demonstrations of the retro-directivity property is that one described in  \cite{AllLeoIto03}, where an antenna array with 8 slot antennas, each of which was connected to a Schottky diode performing heterodyne mixing was realized and tested.  
%
%
Another relatively simple solution is proposed in \cite{PhysRevLett.105.123905}, in which active split-ring resonators loaded with varactor diodes are demonstrated to act as phase-conjugating elements when pumped with a signal at frequency $2f_0$.
In addition to the previous active solutions, also low-complexity passive retro-directive metasurfaces have been recently realized (in this case $g<1$). By properly engineering the surface impedance according to a supercell design periodicity that is greater than the wavelength $\lambda$, it is possible to reflect the incident wave in a direction other than specular. This principle is exploited, for instance, in \cite{KalSee20} to realize a metasurface that exhibits a high level of retro-directivity for several angles of incidence.
%
%


Suppose, now, that the metasurface not only performs the conjugation and (possible) amplification of the received signal  but also introduces, in each $k$th symbol interval, a phase shift $\phi[k]$  (the same for all cells) that incorporates the information to be transmitted  by the \ac{UE} in that interval
(see Fig. \ref{Fig:SCM}). 
As a consequence, the vector of the backscattered signal becomes 
\begin{equation}
\label{eq:SCMmodel_with_info}
\boldr[k]=g \, e^{\jmath \phi[k]} \, \boldz^*[k]  
\end{equation}
where $\boldz[k]=\bolds[k] + \boldeta[k]$.
The conjugation operation will be exploited by the algorithm proposed in the next sections. Since the phase $\phi[k]$ affects all cells of the metasurface,  it does not compromise the retro-directivity behavior as the common phase term can be absorbed by the constant $A$.
Moreover, the implementation of the modulated \ac{SCM} does not require RF/ADC chains as data directly modulates the phase sequence $\{\phi[k] \}$, thus allowing a low-cost, low-complexity, low-energy consumption multi-antenna device.   
%


\begin{figure}[t!]
	\centering \includegraphics[trim= {0 0 0 0}, clip, width=0.9\linewidth]{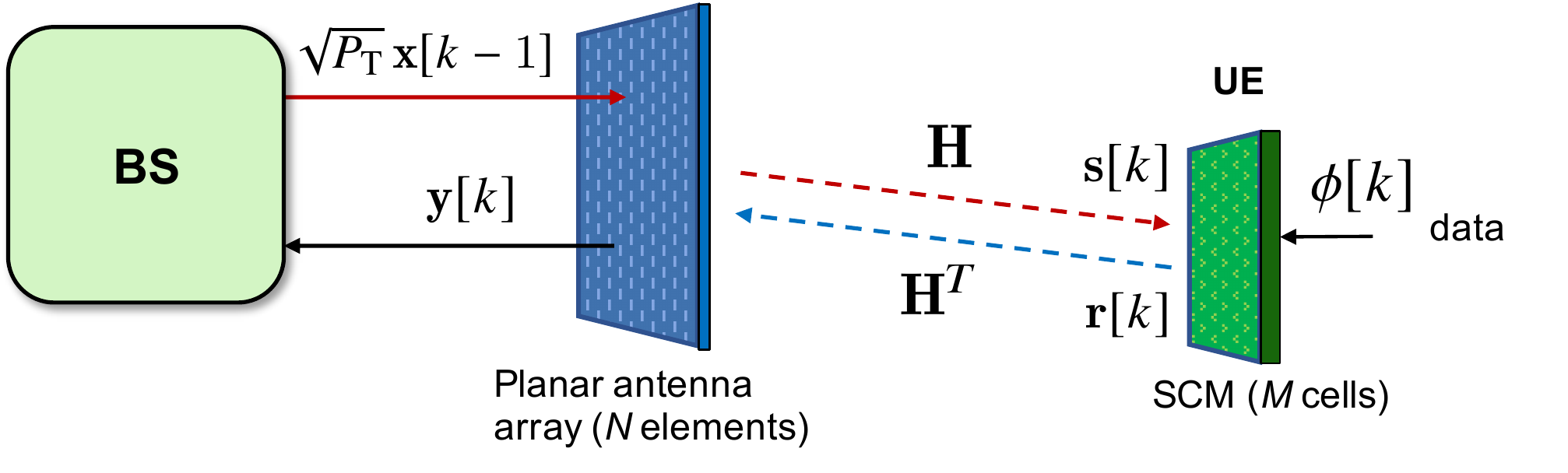}
		\caption{Principle scheme of a SCM-based MIMO communication.}
		\label{Fig:Single}
\end{figure}

\section{Working Principle of SCM-based MIMO Communication}
\label{Sec:Working}

In this section, we start with the single-user uplink scenario, shown in Fig. \ref{Fig:Single},  with the purpose to illustrate how data can be transmitted from an \ac{SCM}-based \ac{UE} to the \ac{BS}.
In particular, we consider a \ac{BS}, equipped with an antenna array of $N$ elements, capable of full-duplex  narrow-band transmission with bandwidth $W$.  The time interval $T$ between time instants $k-1$ and $k$ is in the order of $1/W$.\footnote{When describing the transmitted signal and the signal received/retro-directed by the \ac{UE}, time instants $k$ should be intended as intervals. Sampling is operated at the \ac{BS} after standard matched filter processing of the signal in the last symbol interval of duration $T$.}  
The \ac{UE} is realized according to the scheme described in Sec. \ref{Sec:SCM}, thus incorporating an \ac{SCM}-based antenna  composed of $M$ cells. 
Let $\sqrt{\Ptx}\,\boldx[k-1]\in \mathbb{C}^{N \times 1}$ be the vector containing the signal  transmitted by the $N$ elements of the \ac{BS}'s antenna array during the $k$th time interval $[(k-1)T,k\, T]$, where $\Ptx$ is the transmitted power and $\boldx[k-1]$ is a unitary norm beamforming vector, i.e., the precoding vector designed to steer the beam toward the \ac{UE} according to the scheme proposed in Sec. \ref{sec:RandomAccess}. 
The transmitted signal  is received and retro-directed by the \ac{UE} during the time interval $k$, then collected by the \ac{BS}. 
Here, we  assume  symbol-level synchronization between the \ac{UE} and the \ac{BS}.\footnote{
The analysis of schemes at the \ac{BS} capable of synchronizing with the local clock of the \ac{UE}  is outside the scope of this paper, and it will be the topic of future works.}

The signal received by the \ac{UE} in the $k$th time interval is
\begin{equation}
\boldz[k]=\sqrt{\Ptx}\,  \boldH \, \boldx[k-1] + \boldeta[k]
\end{equation}
where $\boldH \in \mathbb{C}^{M \times N}$ denotes the channel matrix. At the \ac{UE}, information data is associated with the phase sequence $\left \{ \phi[k] \right \}_{k=1}^K$, forming a packet of length $K$ symbols, according to any phase-based signaling scheme (e.g., BPSK).  
Without loss of generality, we consider that all transmitted packets have the same length $K$, even though packets with different lengths can be easily managed by the scheme proposed in this paper without any change. 
The signal backscattered by the UE, according to  \eqref{eq:SCMmodel_with_info}, is
\begin{equation}
\boldr[k]=g \, e^{\jmath \phi[k]} \, \boldz^*[k]
=\, \sqrt{\Ptx}\, g\,  e^{\jmath \phi[k]} \, \boldH^* \, \boldx^*[k-1] + g\, \boldeta^*[k]
\end{equation}
being $\phi[k]$ the phase associated to the $k$th data symbol of the transmitted packet. 
At the \ac{BS} side, the received signal at time interval $k$ is
\begin{equation} \label{eq:byk}
\boldy[k]=\sqrt{\Ptx} \, g \, e^{\jmath \phi[k]} \, \boldH^T \boldH^{*} \boldx^*[k-1] + g\, \boldH^T \, \boldeta^*[k] + \boldw[k] 
\end{equation}  
where $\boldw[k] \sim {\mathcal{CN}}\left (\mathbf{0}, \sigma_w^2 \boldI_N \right )$ is the \ac{AWGN} at the receiver,  $\sigma_w^2=\kappa T_0 \Fbs   \, W$, being $\Fbs$ the \ac{BS}' noise figure. 
The noise term $\boldw[k]$ might also include any clutter backscattered by the surrounding environment whose specific characterization depends on the environment itself. 
 For further convenience, we can rewrite  \eqref{eq:byk} as
\begin{equation} \label{eq:bykbis}
\boldy[k]=\sqrt{\Ptx} \, e^{\jmath \phi[k]} \, \boldA^*\,  \boldx^*[k-1] + \boldn^*[k]
\end{equation}  
where we have defined  $\boldA= g\,  \boldH^{\dag} \boldH \in \mathbb{C}^{N \times N}$, $\boldn^*[k]=g\, \boldH^T \, \boldeta^*[k]+\boldw[k] \sim {\mathcal{CN}}\left (\mathbf{0}, \sigma_w^2 \, \boldI_N+\sigma_{\boldeta}^2   \,  \boldA^* \right )$, which  includes all the noise terms. Note that, because of the conjugation operated by the \ac{SCM},  $\boldA$ is not the round-trip channel which is given, instead, by $g\, \boldH^{T} \boldH$. For this reason, we refer to it as the \emph{modified round-trip channel}.


Denote with $\lambda_j$ the $j$th eigenvalue of $\boldA$, where $\lambda_1\ge \lambda_2 \ge \ldots, \ge \lambda_N$, and $\boldv_j$ is the corresponding eigenvector (direction). 
As a consequence, vector $\boldn[k]$ can be decomposed as
\begin{equation}
\boldn[k]=\sum_{j=1}^{N} n_j[k] \, \boldv_j 
\end{equation}
where $n_j[k]$ is the projection of $\boldn[k]$ onto the $j$th direction $\boldv_j$, and $n_j \sim {\mathcal{CN}}\left (0, \sigma_j^2 \right )$, with $\sigma^2_j=\sigma_w^2+ \lambda_j\, \sigma^2_{\eta}$. 
In a typical practical setting, the modified round-trip channel gain  $\| \boldA \|_{\text{F}}^2$ is much less than one, then the noise component $\boldw[k]$ becomes the dominant term in $\boldn[k]$ and hence we can write with  good approximation 
$\sigma_j^2 = \sigma^2$, $\forall j$, with  $\sigma^2 \simeq\sigma_w^2$ (isotropic noise).
%
 Interestingly, thanks to the conjugating operation at the user's \ac{SCM}, the eigenvectors of $\boldA$ correspond to the left-eigenvectors of the \ac{MIMO} channel $\boldH$, then the estimation of the optimum beamforming vector $\boldx[k]$ is equivalent to computing the top-eigenvector $\boldv_1$ of the modified round-trip channel $\boldA$ at the \ac{BS} without requiring any processing at \ac{UE} side, as it will be evident in Sec. \ref{sec:RandomAccess}.

Information data can be extracted by the \ac{BS} at each time interval by forming the decision variable $u[k]$ through the correlation of  the received vector $\boldy[k]$ with the  complex conjugate of the beamforming vector $\boldx^{\dag}[k-1]$
\begin{equation} \label{eq:uk1}
    u[k]=\boldx[k-1]^{\dag} \, \boldy^*[k] \, .
 \end{equation}
In particular, assuming a perfect estimation of the beamforming vector $\boldv_1$ is available, then $\boldx[k-1]=\boldv_1$ and 
\begin{equation} \label{eq:uk2}
    u[k]=\sqrt{\Ptx} \,  e^{-\jmath \phi[k]}  \, \boldv_1^{\dag} \,  \boldA\,  \boldv_1 + \boldv_1^{\dag} \,\boldn[k]=\sqrt{\Ptx} \, \lambda_1 \, e^{-\jmath \phi[k]} + n_1[k] 
\end{equation}
with $n_1 \sim {\mathcal{CN}}\left (0, \sigma^2 \right )$. 
Starting from $u[k]$, a decision on the phase $\phi[k]$ can be made by taking its argument, i.e.,   
$\hat{\phi}[k]=\mathsf{demodulation}(-\arg u[k])$, where $\mathsf{demodulation}(\cdot)$ is the data detection function which depends on the particular signaling scheme considered. Without loss of generality, in our numerical results, we will consider the BPSK scheme.
The corresponding \ac{SNR} at the decision device is 
\begin{equation} \label{eq:SNR0}
\mathsf{SNR}=\frac{\Ptx \, \lambda_1^2}{\sigma^2} \, .
\end{equation}

It is worth noticing that when the channel has rank ${r=1}$, the resulting communication scheme is capacity-optimal provided that an accurate estimation of the beamforming vector $\boldv_1$ is available.  When $r>1$, the scheme is not capacity-optimal because the \ac{SCM} is intrinsically single-layer so that only one out of the $r$ potential data streams,  which could be established between the \ac{BS} and the \ac{UE}, is exploited.
However, it corresponds to the optimal single-layer beamforming scheme in the \ac{SNR} maximization sense also providing the maximum diversity gain \cite{LarStoB:03}.  

It is interesting to analyze the \ac{SNR} in free-space condition assuming  the \ac{BS} and the \ac{SCM} are in paraxial configuration at distance $d$. 
In this case, the  \ac{SNR} in \eqref{eq:SNR0} becomes 
%
\begin{align} \label{eq:SNRfree}
    \mathsf{SNR}&=\frac{\Ptx\, g^2 \, N^2\, M^2\, \Gbs^2 \, \Gscm^2 \,  \lambda^4}{ \sigma^2 \, \left ( 4 \pi \, d \right )^4}  \
\end{align}
where $\Gbs$ and  $\Gscm$ are the gain of the elements of the \ac{BS}'s antenna and the \ac{SCM}'s cell, respectively. 
The last equation shows that, due to the backscattering nature of the communication, the path loss increases with the distance to the power of four, as happens with \ac{RFID} systems based on backscatter modulation \cite{DarDErRobSibWin:J10}. On the other hand, such  large path loss can be easily compensated by increasing the number of antenna elements $N$ and $M$ at the \ac{BS} and \ac{UE}, respectively. 

Now, in order to establish the  single-layer optimal communication between the \ac{UE} and the \ac{BS}, the main problem to be solved  is the fast detection of a new user and the fast and accurate  estimation of the beamforming vector  $\boldv_1$, especially in a multi-user \ac{MIMO}  scenario.
In the next section, a fast and blind (i.e., without \ac{CSI} estimation) scheme for joint eigenvectors estimation and multi-user communication is presented.

\begin{figure}[t!]
	\centering \includegraphics[trim= {0 0 0 0}, clip, width=1\linewidth]{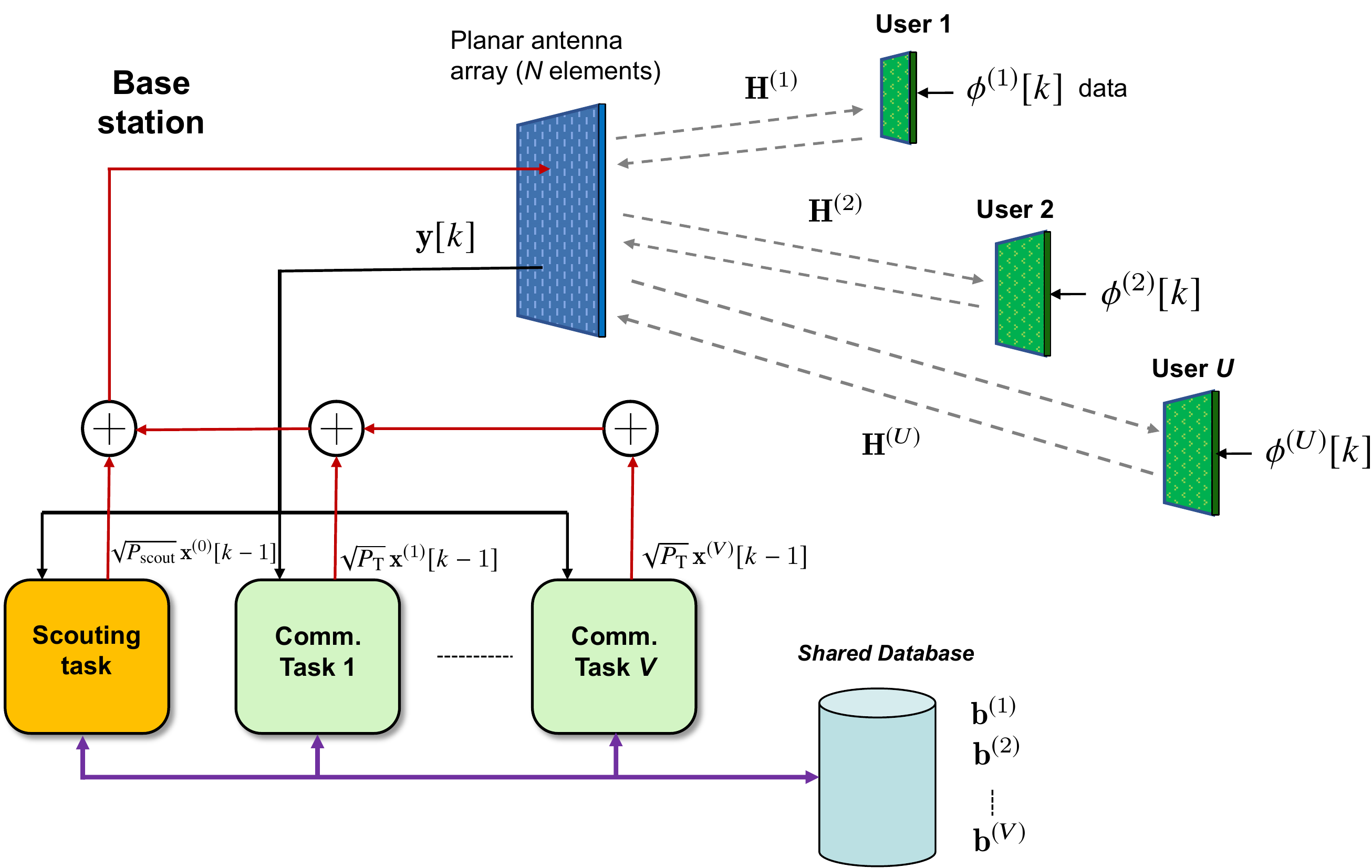}
		\caption{Scheme of the proposed grant-free \ac{MIMO} access based on \acp{SCM}.}
		\label{Fig:Scheme}
\end{figure}

\section{Grant-free Random Access using MIMO SCMs}
\label{sec:RandomAccess}

Consider  a scenario in which a large number of asynchronous \acp{UE} deployed in the environment generate packets randomly.
We consider the worst-case situation where each packet is generated by a distinct \ac{UE} located at a different  location so that the reception of each packet requires the fast detection of the new \ac{UE} and the fast estimation of the beamforming vector at the \ac{BS} without any a priori information. For this reason, in what follows we will interchange freely the terms packets and \acp{UE}.  Note that in such a scenario, all the approaches relying on statistical/long-term \ac{CSI} and/or \acp{UE} activity pattern estimation are highly inefficient or not applicable (e.g., \cite{BanXuCarPop:18}). 

Suppose that, at the generic time interval, there are $U$ asynchronous active (i.e., transmitting) \acp{UE}. Denote with $\mathbf{H}^{(1)}$, $\mathbf{H}^{(2)}$, ... ,$\mathbf{H}^{(U)} \in \mathbb{C}^{M \times N}$ the channel matrices related to the $U$ links between the \ac{BS} and the active \acp{UE}. For further convenience, we define the corresponding modified round-trip channels $\boldA^{(u)}= g \, \left (\boldH^{(u)}\right )^{\dag}  \boldH^{(u)} \in \mathbb{C}^{N \times N}$, for $u=1,2, \ldots, U$. In addition, we denote with $\boldv_i^{(u)}$ and $\lambda_i^{(u)}$, respectively, the $i$th eigenvector and eigenvalue of $\boldA^{(u)}$, with ${i=1, 2, \ldots , r^{(u)}}$, where $r^{(u)}=\rank{\boldA^{(u)}}$.

The main idea of the proposed grant-free \ac{MIMO} access scheme is sketched in Fig. \ref{Fig:Scheme} and described in the following. 
The \ac{BS} has several  tasks running in parallel. In particular, one task, namely \emph{Scouting Task}, is always active and its purpose is to discover any newly transmitted packet from a new active \acp{UE} and estimate the corresponding, possibly optimal, beamforming vector at the \ac{BS}.   Once a new packet has been detected, the estimated beamforming vector, namely $\boldb^{(v)}$, is added to a \emph{Shared Database}, containing all the estimated beamforming vectors of the packets currently under decoding. 
In addition, a dedicated task, denoted to as  \emph{Communication Task},  is instantiated and associated to the new packet in charge of decoding it until its end. When the packet is over,  i.e., the \ac{UE} stops transmitting, the associated Communication Task is terminated and the corresponding beamforming vector is removed from the Shared Database.  As it will be explained later, if $r^{(u)}>1$, then the same packet transmitted by the $u$th \ac{UE} might be decoded using different orthogonal beamforming vectors, each of them associated with a different eigenvalue of $\boldA^{(u)}$. As a consequence, the size of the Shared Database, and then the number $V$ of instantiated Communication Tasks, might be in general equal or larger than the number $U$ of currently active \acp{UE}, that is,  $V\ge U$.
Denote with $u=f(v)$ the function that maps the packet under decoding by the $v$th Communication Task with the index $u$ of the \ac{UE}  that generated it.    

According to the scheme  in Fig. \ref{Fig:Scheme}, the Scouting Task and the $V$ Communication Tasks generate, respectively,  the beamforming vectors $\sqrt{\Pscout} \, \boldx^{(0)}[k-1]$ and $\sqrt{P^{(v)}} \, \boldx^{(v)}[k-1]=\sqrt{P^{(v)}} \, \boldb^{(v)}$, $v=1,2, \ldots V$,  which are summed up to form the actual transmitted vector $\boldx[k-1]$ at time $k-1$, that is, 
\begin{equation}
\boldx[k-1]= \sum_{u=0}^V  \sqrt{P^{(v)}} \, \boldx^{(v)}[k-1]   
\end{equation}
where $P^{(0)}=\Pscout$ and $\boldx^{(0)}[k-1] $ are, respectively,  the transmitted power  and the current beamforming vector used by the Scouting Task. 
Without loss of generality, we assume that all the Communication Tasks transmit with the same  power $\Ptx$, so  $P^{(v)}=\Ptx$, $v=1,2, \ldots V$.
The signal received by the \ac{SCM} of the  $u$th \ac{UE} during the $k$th transmission interval  is
\begin{equation} \label{eq:boldzk}
    \boldz^{(u)}[k]= \boldH^{(u)} \, \boldx[k-1]+ \boldeta^{(u)}[k]= \boldH^{(u)} \, \sum_{l=0}^V \sqrt{P^{(l)}} \,  \boldx^{(l)}[k-1] + \boldeta^{(u)}[k]
\end{equation}
where ${\boldeta^{(u)} [k]\sim {\mathcal{CN}}\left (\mathbf{0}, \sigma_{\eta}^2\,  \boldI_M \right )}$ is the noise of the  \ac{SCM} of the $u$th \ac{UE}. In \eqref{eq:boldzk}, $\boldH^{(0)}$ represents the channel matrix between the \ac{BS} and the possibly new \ac{UE} under scouting. Therefore $u \in \cal{U}$, where ${\cal{U}}=\{ 0, 1, 2, \ldots , U \}$, if a new \ac{UE} is present, whereas ${\cal{U}}=\{ 1, 2, \ldots , U \}$ otherwise.
According to \eqref{eq:SCMmodel_with_info}, the signal backscattered by the $u$th \ac{UE} is  
$\boldr^{(u)}[k]=g \, e^{\jmath \phi^{(u)}[k]} \left (\boldz^{(u)}[k] \right )^*$, then the signal received by the \ac{BS} at the $k$th time interval is given by the contributions from all \acp{UE} and can be written as
\begin{align} \label{eq:boldylbis}
    \boldy[k]
    &=\sum_{u \in \cal{U}}   \left (\boldH^{(u)} \right )^T \, \boldr^{(u)}[k] +\boldw[k] \nonumber \\
    &=\sum_{u \in \cal{U}} e^{\jmath \phi^{(u)}[k]} \left (\boldH^{(u)} \right )^T  \left (\boldH^{(u)} \right )^* \sum_{l=0}^V g\, \sqrt{P^{(l)}} \,   \, \left (\boldx^{(l)}[k-1]\right )^* + \boldn[k]  \nonumber \\
    &=\sum_{u \in \cal{U}} e^{\jmath \phi^{(u)}[k]}  \left (\boldA^{(u)} \right )^* \sum_{l=0}^V  \sqrt{P^{(l)}} \,   \, \left (\boldx^{(l)}[k-1]\right )^* + \boldn[k]   
\end{align}
where $\boldA^{(0)}=g\, \left (\boldH^{(0)} \right )^{\dag} \boldH^{(0)}$, 
$\boldn[k] = \boldw[k] + \sum_{u \in \cal{U}} \left (\boldH^{(u)} \right )^T  \left (\boldeta^{(u)}[k] \right )^* $ 
and $\left \{  \phi^{(u)}[k]\right \}$ is the data packet of the $u$th active \ac{UE}.
All the tasks  have access to the received vector $\boldy[k]$.

\begin{algorithm}[t]
\footnotesize
\setstretch{1.1}
\DontPrintSemicolon
\SetAlgoLined
\textit{Initialization}: generate a guess unitary norm beamforming vector $\boldx^{(0)}[0]$; $V=0$;\\
  \For{$k=1,\dots, \infty$}  {
   If the Shared Database has been updated, then generate a random unitary norm vector  $\mathbf{x}^{(0)}[k-1]$ orthogonal to $\boldB$ \tcp*{}
  transmit: $\sqrt{\Pscout} \, \boldx^{(0)}[k-1]$ \tcp*{}
  receive: $\boldy[k]$    \tcp*{}
  $\oy[k]=\mathsf{orthogonalize}(\boldy[k],\boldB) $  \tcp*{}
  $\boldx^{(0)}[k]={\left (\oy[k] \right )^*}/{\left\| \oy[k]\right\|}$ \tcp*{beamforming vector update}
  $\gamma[k]=\left \|  \oy[k] \right \|^2/\sigma^2$ \tcp*{SNR computation}
  If $\gamma[k]>\gamma_{\text{dec}}$ and $\gamma[k]/\gamma[k-1]<\gamma_{\Delta}$,  then a new packet has been detected,  $V=V+1$, a new dedicated Communication Task is allocated using $\boldx^{(0)}[k]$ as beamforming vector, and $\boldb^{(V)}=\boldx^{(0)}[k]$ is added to the Shared Database. 
  }
  \caption{Algorithm of the Scouting Task.}
  \label{Algorithm_1}
\end{algorithm}

The scope of the Shared Database, containing the all the estimated beamforming vectors\\ ${\boldB= \left \{ \boldb^{(1)}, \boldb^{(2)}, \ldots , \boldb^{(V)} \right \} }$ of the packets currently under decoding, is two-fold:  (i) the $v$th beamforming vector $\boldb^{(v)}$ is used by the $v$th Communication Task to address the \ac{UE} with index $u=f(v)$ until the end of the corresponding packet; (ii) the set of beamforming vectors  in $\boldB$ is used by the Scouting Task to perform the estimation of the beamforming vector of a possibly newly transmitting \ac{UE} in a subspace orthogonal to that spanned by the vectors in $\boldB$ (i.e., in the null space of $\boldB$). The latter ensures that the vectors in $\boldB$ are orthogonal and hence all the tasks (Scouting and Communication) address orthogonal subspaces.   

\subsection{Scouting Task}
The pseudo-code of the algorithm running on the Scouting Task is reported in Algorithm \ref{Algorithm_1}.
 At the startup time $k=0$, $\boldx^{(0)}[0]$ is initialized with a unitary norm random vector.  
At the $k-1$th iteration, with $k>1$, the Scouting Task Algorithm (see \ref{Algorithm_1}) sends the current estimated beamforming vector, $\boldx^{(0)}[k-1]$, and receives back the response at time interval $k$ from the new \ac{UE}  (if any) according to \eqref{eq:boldylbis}. Before further processing, the received vector $\boldy[k]$ is made orthogonal to $\boldB$, that is, 
\begin{equation} \label{eq:othogonal}
    \oy[k]
    =\left ( \boldI - \boldB \, \boldB^{\dag} \right )  \boldy[k]  
\end{equation} 
in order to force the searching in the null space of $\boldB$ and hence not  interfere with the currently active Communication Tasks. 
Then, a normalized and conjugated version of $ \oy[k]$ is computed and used as the updated beamforming vector $\boldx^{(0)}[k]$ in the subsequent iteration. 
The processing operated by Algorithm~\ref{Algorithm_1}   is the modification of the well-known \emph{Power Method} \cite{GolVor20} that, in the absence of noise, interference,  and the orthogonalization in \eqref{eq:othogonal},  allows the estimation of the top-eigenvector  of the square matrix $\boldA=\sum_{u \in \cal{U}} \boldA^{(u)}$. 
Here, differently from the classical Power Method, due to \eqref{eq:othogonal}, the matrix seen by the iterative method is $\oA=\left (\boldI - \boldB \, \boldB^{\dag} \right ) \boldA$, which determines the  subspace under exploration  by the Scouting Task,  whose eigenvectors and eigenvalues are denoted by $\left \{ \olambda_1, \olambda_2, \ldots , \olambda_{\orank} \right \}$ and $\left \{ \ov_1, \ov_2, \ldots , \ov_{\orank} \right \}$, respectively, where $\orank=\rank{\oA}$. 
Moreover, every time a change is operated in  $\boldB$ (addition or removal of elements), $\boldx^{(0)}[k]$ is reinitiated by drawing a random  vector subsequently made orthogonal to the vectors in $\boldB$, as in \eqref{eq:othogonal}, and unitary. 
%
 In particular, the beamforming vector is updated according to
\begin{equation} \label{eq:PowerMethod}
    \boldx^{(0)}[k]=\frac{ \oy^*[k]}{\left \| \oy[k] \right \|}
\end{equation}
until a change on $\boldB$ occurs or a new packet is detected.
In the absence of noise and interference, after a few iterations, the Power Method ensures that the direction of $\boldx^{(0)}[k]$ converges to that of the dominant eigenvector $\ov_1$ of $\oA$  \cite{GolVor20}. The convergence behavior of  \eqref{eq:PowerMethod} in the presence of noise 
will be analyzed later.

Obviously, the new \ac{UE} can be detected only if $\oA$ and $\boldA^{(0)}$ share a not null subspace. 
%
%
It has to be remarked that, in general, it is not guaranteed that $\ov_1$ is referred to as the channel between the \ac{BS} and the new \ac{UE}.
In fact, when the \ac{UE} and \ac{BS} are in radiating near-field conditions, or in far-field in the presence of multipath, the channel might have  a rank larger than one and the same \ac{UE} can be reached addressing different orthogonal subspaces \cite{DecDar:J21}. 
In such cases, the Scouting Task is likely to converge to a subspace related to  an \ac{UE} already under decoding but not yet  included in $\boldB$, and hence a second Communication Task  will be associated with that \ac{UE}. 
Duplicated packet decoding is not an issue as it can be easily solved by  higher protocol layers.  On the contrary, such a phenomenon  might be beneficial as it creates a sort of eigenvector diversity. 
If  the packet generation of the new \ac{UE}  is not simultaneous (at symbol time $T$ level) to the generation of an already serving \ac{UE}, 
 it is likely that all the significative secondary eigenvectors of  $\oA$ have been already removed when the new \ac{UE} starts its transmission,  and $\boldv_1^{(0)}$ is effectively associated with the new \ac{UE}, as we will assume in the rest of the paper.

Finally, denote with $k^*$ the time instant a new \ac{UE} starts the transmission of a packet. The new packet is detected when the \ac{SNR} $\gamma[k]=\left \|  \oy[k] \right \|^2/\sigma^2$, with $k>k^*$,   satisfies $\gamma[k]>\gamma_{\text{dec}}$ and $\gamma[k]/\gamma[k-1]<\gamma_{\Delta}$. The first inequality checks that a certain level of energy has been received compared to the noise level, whereas the second one guarantees that \eqref{eq:PowerMethod} is close to the convergence, with the thresholds $\gamma_{\text{dec}}$ and $\gamma_{\Delta}$ to be properly tuned.   

\subsection{Communication Task}
Once a new packet has been detected by the Scouting Task, a new Communication Task is activated whose algorithm is reported in Algorithm  \ref{Algorithm_2}.  
Denote with $k^{(v)}$ the time instant the packet is detected by the Scouting Task and the Communication Tasks starts, being $v$ the index associated to the new packet generated by the $u$th \ac{UE}, with $u=f(v)$, and with $\boldb^{(v)}=\boldx^{(0)}\left [k^{(v)} \right ]$ the corresponding beamforming vector estimated through Algorithm \ref{Algorithm_1} and stored in the Shared Database by the Scouting Task.
The Communication Task, for $k\ge  k^{(v)}$ until the end of the packet, executes Algorithm \ref{Algorithm_2} which, at each iteration, always transmits  the same beamforming vector  $\sqrt{P^{(v)}} \,\boldx^{(v)}[k-1]$, with $\boldx^{(v)}[k-1]=\boldb^{(v)}$. 
Subsequently, according to \eqref{eq:uk1} and \eqref{eq:boldylbis}, the decision variable is formed as  
\begin{align}  \label{eq:uuk}
    u^{(v)}[k]=&\left ( \boldb^{(v)} \right )^{\dag} \boldy^*[k] \nonumber \\
    =& \sum_{i \in \cal{U}} e^{-\jmath \phi^{(i)}[k]}  \, \left ( \boldb^{(v)} \right )^{\dag} \boldA^{(u)} \sum_{l=0}^V \sqrt{P^{(l)}} \,    \boldx^{(l)}[k-1]  + \left ( \boldb^{(v)} \right )^{\dag} \, \boldn^*[k]  \nonumber \\
    =&   e^{-\jmath \phi^{(u)}[k]}  \, \sum_{l=0}^V \sqrt{P^{(l)}} \,\left ( \boldb^{(v)} \right )^{\dag} \boldA^{(u)}   \,   \boldx^{(l)}[k-1] \nonumber \\
    & +  \sum_{{i \in \cal{U}}, \, i\neq u}^U e^{-\jmath \phi^{(i)}[k]}  \, \sum_{l=0}^V \sqrt{P^{(l)}} \,   \left ( \boldb^{(v)} \right )^{\dag} \boldA^{(u)}  \,  \boldx^{(l)}[k-1]  + \left ( \boldb^{(u)} \right )^{\dag} \, \boldn^*[k]  \, .
\end{align}
%
%
A decision on the transmitted phase $\phi^{(u)}[k]$ from the \ac{UE} with index $u=f(v)$ is made based on the argument  of $u^{(v)}[k]$ in \eqref{eq:uuk} according to the mapping scheme considered, as explained in Sec.  \ref{Sec:Working}. 
The first term in  \eqref{eq:uuk}  is the useful term. 
Equation \eqref{eq:uuk} reveals that, in general, the decision variable is affected by the presence of interference from other \acp{UE}  (second term), caused by eventual not orthogonality of the users' channels,  and thermal noise (third term).  
Finally, when $\gamma[k]=\left | u^{(v)}[k] \right |^2/\sigma^2<\gamma_{\text{drop}}$, the end of the packet has been likely reached, and the task is deactivated. 

\begin{algorithm}[t]
\footnotesize
\setstretch{1.1}
\DontPrintSemicolon
\SetAlgoLined
\textit{Initialization}: $active=1$, $k=k^{(v)}$  \tcp*{ $k^{(v)}$ time instant of packet detection} 
   \While{active=1} {
   transmit: $\sqrt{P^{(v)}} \, \boldx^{(v)}[k-1]=\sqrt{P^{(v)}} \,  \boldb^{(v)}$ \tcp*{}
  receive: $\boldy[k]$    \tcp*{}
  $u^{(v)}[k]= \left (\boldb^{(v)} \right )^{\dag} \, \boldy^*[k]  $ \tcp*{decision variable}
  $\widehat{\phi}[k]= \mathsf{demodulation} \left (-\arg \left \{u^{(v)}[k]  \right \} \right ) $ \tcp*{data demodulation}
  $\gamma[k]=\left |   u^{(v)}[k] \right |^2/\sigma^2$ \tcp*{SNR computation}
  If $\gamma[k]<\gamma_{\text{drop}}$, then beamforming vector $\boldb^{(v)}$ is removed from the Shared Database and this task is de-instantiated ($active=0$, $V=V-1$) \tcp*{}
  $k=k+1$; 
  }
  \caption{Algorithm of the $v$th Communication Task.}
  \label{Algorithm_2}
\end{algorithm}

\subsection{Orthogonal Channels (favorable propagation)}

Equation \eqref{eq:uuk} shows that,  depending on channel characteristics, the decision variable might be deteriorated by the presence of inter-user interference. 
In this section, we investigate  the  particular but significative scenario where the favorable propagation condition is met, which is typical when employing a massive number of antenna elements at the \ac{BS} \cite{SanguinettiBook:2017}. 

Consider the scenario in which the subspaces spanned by $\mathbf{H}^{(u)}$ are orthogonal to each other,
and so those by $\boldA^{(u)}$, for $u \in \cal{U}$. This implies that 
$\left ( \boldv_n^{(v)} \right )^{\dag}  \boldA^{(u)} \, \boldv_m^{(l)}$ is different from zero only when $v=l$ and  $m=n$.  
This assumption can be in general approached when \acp{UE} are located in different positions and a large number of antennas is used at the \ac{BS}. This condition is typically referred to as \emph{favorable propagation} in massive \ac{MIMO} literature.
For instance, in \cite{WuLiu:17} it has been shown that the favorable propagation condition can be approached even in correlated Rayleigh and geometry-based multi-user \ac{MIMO}  channels when $N$ grows.

Under favorable propagation and  supposing an accurate estimate of the eigenvectors has been reached by the Scouting Task so that $\boldb^{(v)} \simeq \boldv_i^{(u)}$, with $u=f(v)$, for some $i \in \{1,2, \ldots, r^{(u)}\}$,   the inter-user interference terms become negligible and \eqref{eq:uuk} simplifies to  
\begin{align} \label{eq:uk0}
    \boldu^{(v)}[k]    \simeq & \sqrt{\Ptx} \,  e^{-\jmath \phi^{(u)}[k]} \left ( \boldb^{(v)} \right )^{\dag} \boldA^{(u)} \,  \boldb^{(v)}   +  \left ( \boldb^{(v)}\right )^{\dag} \, \boldn^*[k]  
    \simeq   \sqrt{\Ptx} \lambda_i^{(u)} e^{-\jmath \phi^{(u)}[k]}   + n_i[k]  
\end{align}
like in the single-user case in \eqref{eq:uk2} because $\boldb^{(v)}$ is an eigenvector of  the round-trip time channel matrix $\boldA^{(u)}$. In addition, $\oA=\boldA^{(0)}$ and $\ov_1=\boldv_1^{(0)}$, that is, the top-eigenvector of $\boldA^{(0)}$.

\section{Convergence Analysis of the Scouting Task}
\label{sec:Convergence}


The convergence behavior of the Scouting Task, given a new \ac{UE} started its transmission, can be analyzed assuming that  no other \ac{UE} starts its transmission during the first symbols of the packet and/or has its channel subspace partially overlapped in the subspace spanned by $\oA$. 
%
In the presence of noise, \eqref{eq:PowerMethod} reads
\begin{equation}
    \boldx^{(0)}[k]=\frac{\oy^*[k] }{\left\|\oy[k]\right\|}=\frac{ \sqrt{\Pscout} \,  \oA \, e^{-\jmath \phi[k]} \boldx^{(0)}[k-1] + \boldn[k]}{ \left\| \sqrt{\Pscout} \, \oA \, e^{-\jmath \phi[k]} \boldx^{(0)}[k-1] + \boldn[k] \right\|}
\end{equation}
where 
\begin{equation}\label{eq:ywithnoise}
    \oy^*[k] =\sum_{j=1}^N \ov_j \left (\sqrt{\Pscout} \, \olambda_j x_j[k-1] e^{-\jmath \phi[k]}  + n_j[k] \right ) 
\end{equation}
 and $x_j[k]$, $n_j[k] \sim {\mathcal{CN}}\left (0, \sigma^2  \right )$ represent, respectively, the projections of $\boldx^{(0)}[k]$, $\boldn[k]$ onto $\ov_j$. 
Note that the term carrying the information (i.e., the phase $\phi[k]$) also includes the noise from the previous iterations, which is contained in $x_j[k-1]$.
Moreover, $\left |x_j[k-1]\right |^2 / \left \| \boldx^{(0)}[k-1] \right \|^2=|x_j[k-1]|^2$, being $\left\|\boldx^{(0)}[k]\right\|^2=1$, represents  the fraction of the total transmitted power (useful plus noise) associated to direction $\ov_j$ at the discrete time $k-1$. Then, at the end of the $k$th time interval, the received \ac{SNR}  along the direction $\ov_j$ is given by
 \begin{align}
 \mathsf{SNR}_j[k]=\frac{\Pscout \, \olambda_j^2 \, |x_j[k-1]|^2}{\sigma^2} \, .
 \label{eq:SNRj2}
 \end{align}

The goal is to determine an iterative expression for $\mathsf{SNR}_j[k]$, which drives the signal demodulation performance in the Communication Task, and evaluate the convergence condition of the Scouting Task we proposed.
Considering \eqref{eq:ywithnoise}, the fraction of the total power that is associated with direction $\ov_j$ at the beginning of time interval $k$  can be written as
\begin{align}
|x_j[k]|^2=\frac{\Pscout \, \olambda_j^2 \, |x_j[k-1]|^2 + \sigma^2}{\sum_{i=1}^N \left ( \Pscout \, \olambda_i^2 \, |x_i[k-1]|^2  + \sigma^2 \right )} \, .
\label{eq:etaj_def}
\end{align}
By inverting \eqref{eq:SNRj2} and plugging $|x_j[k]|^2$ at both the left-hand and right-hand sides of \eqref{eq:etaj_def}, we obtain the following iterative formula for $\mathsf{SNR}_j[k]$ 
\begin{align}
\mathsf{SNR}_j[k]=&  \frac{\Pscout \,\olambda_j^2\left(\mathsf{SNR}_j[k-1]+1\right)}{\sigma^2\left[\sum_{i=1}^N  \left ( \mathsf{SNR}_i[k-1] +1\right )\right]} \nonumber \\
= &\mathsf{SNR}_j^{(\mathrm{max})}  \frac{\mathsf{SNR}_j[k-1]+1}{N+ \sum_{i=1}^{\orank} \mathsf{SNR}_i[k-1]}
\label{eq:SNRj1}
\end{align}
for $j=1,2, \ldots , \orank$, and $k \ge k^{*}$,  where $\mathsf{SNR}_j[k^*]=\mathsf{SNR}_j^{(\mathrm{max})} \left |x_j[k^*]\right |^2$, and 
\begin{align}
\mathsf{SNR}_j^{(\mathrm{max})}=\frac{\Pscout \, \olambda_j^2 }{\sigma^2} 
 \label{eq:SNRj}
 \end{align}
 representing the  maximum \ac{SNR} along the direction $\ov_j$, i.e., the \ac{SNR} one would obtain if all the power was concentrated to direction $\ov_j$.  Note that if $\boldx \left [k^* \right ]$ is initialized randomly, then  $\left |x_j[k^*]\right |^2 \simeq 1/N$ so that $\mathsf{SNR}_j[k^*] \simeq \mathsf{SNR}_j^{(\mathrm{max})}/N$.
 
 In the particular but common case where channel $\oA$ has rank 1, 
the \ac{SNR}  at the $k$th time instant along direction $\ov_1$ in \eqref{eq:SNRj1} simplifies into 
\begin{align}
\mathsf{SNR}_1[k]=\mathsf{SNR}_1^{(\mathrm{max})}  \frac{\mathsf{SNR}_1[k-1]+1}{N+ \mathsf{SNR}_1[k-1]}  
\label{eq:SNR1}
\end{align}
for $k\ge k^*$, which allows an easy evaluation of the convergence value.   At the convergence it must be $\mathsf{SNR}_1[k] \simeq \mathsf{SNR}_1[k-1]$  then, by dropping the time index,  \eqref{eq:SNR1} becomes the quadratic equation  
\begin{equation}
\mathsf{SNR}_1=\mathsf{SNR}_1^{(\mathrm{max})}  \frac{\mathsf{SNR}_1+1}{N+ \mathsf{SNR}_1}  
\end{equation} 
whose positive solution is 
\begin{equation}
\label{eq:sol}
\mathsf{SNR}_1=\frac{1}{2} \left ( N-\mathsf{SNR}_1^{(\mathrm{max})} + \left |\mathsf{SNR}_1^{(\mathrm{max})}-N \right | \sqrt{1+\frac{\mathsf{SNR}_1^{(\mathrm{max})}}{(\mathsf{SNR}_1^{(\mathrm{max})}-N)^2}}  \right ) \, .
\end{equation}

Denote with $\mathsf{SNR}_1^{(\mathrm{boot})}=\mathsf{SNR}_1^{(\mathrm{max})}/N$ the  \emph{bootstrap \ac{SNR}}. 
By inspection of \eqref{eq:sol} it can easily verified that if  $\mathsf{SNR}_1^{(\mathrm{boot})}\gg1$, then at the convergence it is $\mathsf{SNR}_1 \simeq \mathsf{SNR}_1^{(\mathrm{max})}-N \simeq \mathsf{SNR}_1^{(\mathrm{max})} \gg 1$, 
which takes the role of asymptotic \ac{SNR} corresponding to the optimum beamforming vector for a channel with rank 1. 
When the bootstrap \ac{SNR} is much less than one, we still have convergence but at 
$\mathsf{SNR}_1 \ll 1$ 
so that the link cannot be established. 
It is worth noticing that the convergence value does not depend  on the initial random guess,  but only on the bootstrap \ac{SNR}. 
According to \eqref{eq:SNRfree}, the link budget can be ameliorated by increasing $M$ and/or $N$ indifferently as both play quadratically. On the contrary, the bootstrap \ac{SNR} increases quadratically with $M$ but only linearly with $N$, therefore it is better to increase $M$ than $N$, when possible. 
%
%
The convergence analysis for $\orank>1$ involves the evaluation of an $\orank +1$ polynomial equation that becomes not feasible analytically for large $r$. Despite that, numerical investigations reported in the next section put in evidence that the previous condition to reach the asymptotic \ac{SNR}  holds even when $\orank>1$ so that, upon convergence,  $|x_1[k
]|\gg |x_j[k]|$ for $j>1$.

%

 When the Scouting Task detects the new \ac{UE},  let's say at time interval $\tilde{k}$, $\boldx^{(0)} \left [\tilde{k} \right ]$ is used as an estimate of the top-eigenvector $\ov_1$ in the Communication Task that will be associated to the \ac{UE}. Such an estimate is characterized by  $\mathsf{SNR}_1 \left [\tilde{k}\right ]$ that can be computed through the iterative formula \eqref{eq:SNRj1}.  
The decision variable $u[k]$ at the $k$th symbol  is proportional to the product
\begin{align}
    u[k] \propto  & \, \boldx^{\dag}[\tilde{k}] \, \boldy^*[k]
    =  \sqrt{\Ptx} e^{-\jmath \phi[k]} \, \sum_{j=1}^{\orank} \olambda_j \left|x_j[\tilde{k}]\right|^2  + \boldx^{\dag}[\tilde{k}] \, \boldn[k]  
\end{align}
in which the first term is the useful one, as it contains the phase $\phi[k]$ that carries the information, and the second term represents the noise. 
%
%
%
Since $\left\|\boldx[\tilde{k}]\right\|^2=1$, the \ac{SNR} at the input of the detector 
is 
\begin{align}  \label{eq:SNRdec1}
\mathsf{SNR}^{(\mathrm{dec})}=\frac{\Ptx \left (  \sum_{j=1}^{\orank} \olambda_j \, \left|x_j[\tilde{k}]\right|^2 \right )^2}{\sigma^2} \, .
\end{align}
Therefore, we can rewrite \eqref{eq:SNRdec1} as a function of $\mathsf{SNR}_j[k]$ as
%
\begin{align} \label{eq:SNRdecj}
\mathsf{SNR}^{(\mathrm{dec})}= \frac{\Ptx}{\Pscout} \left ( \sum_{j=1}^{\orank}  \frac{\mathsf{SNR}_j[\tilde{k}]}{\sqrt{\mathsf{SNR}_j^{(\mathrm{max})}}}  \right )^2  \simeq 
 \frac{\Ptx}{\Pscout} \mathsf{SNR}_1^{(\mathrm{max})} =\frac{\Ptx \, \olambda_1^2 }{\sigma^2} \, .
\end{align}

In case of  favorable propagation it is 
\begin{align}
\mathsf{SNR}^{(\mathrm{dec})} \simeq \frac{\Ptx \,\left ( \lambda_1^{(0)} \right )^2 }{\sigma^2} 
  \end{align}
which resembles \eqref{eq:SNR0} of the single-user scenario.

{
\begin{table}[t]
\caption{Parameters used in the analysis and the simulation}
\begin{center}
\footnotesize
\begin{tabular}{|l|l|l|}
\hline
Parameter & Symbol & Value\\
\hline
Carrier frequency & $\fc$  & 100 GHz  \\
BS antenna element gain & $\Gbs$ & 0 dB (isotropic) \\
SCM single cell gain & $\Gscm$ & 0 dB (isotropic) \\
SCM backscatter gain & $g$ & 20 dB \\
Bandwidth & $W$ & 10 MHz\\
Symbol time & $T$ & $100\,$ns\\
TX power  & $\Ptx$ & -5 dBm\\
Power boost scouting & $\Pscout/\Ptx$ & $10$ dB\\  
SCM cell noise figure & $\Fscm$ & 3 dB\\
BS noise figure & $\Fbs$ & 3 dB\\
BS antenna elements & $N$ & $30 \times 30$ ($4.5\times4.5\,$cm$^2$ planar array)\\
SCM cells per user  & $M$ & $20 \times 20$ ($3\times3\,$cm$^2$ planar array)\\
Path-loss exponent & $\beta$ & 2 (free-space/LOS), 2.5 (NLOS)\\
Detection SNR threshold & $\gamma_{\text{dec}}$ &  30 dB \\
Drop SNR threshold & $\gamma_{\text{drop}}$ &  5 dB\\
Delta SNR convergence & $\gamma_{\Delta}$ &  5 dB\\
Packet length & $K$ & 144 bits\\
Guard symbols & $K_{\text{g}}$ & 16 bits\\
Packet duration & $T_{\text{p}}=K\, T$ & $14.4\,\mu$s\\
\hline 
\end{tabular}
\end{center}
\label{Tab:Parameters}
\end{table}

\begin{figure}[t!]
	\centering \includegraphics[trim= {0 0 0 0}, clip, width=0.8\linewidth]{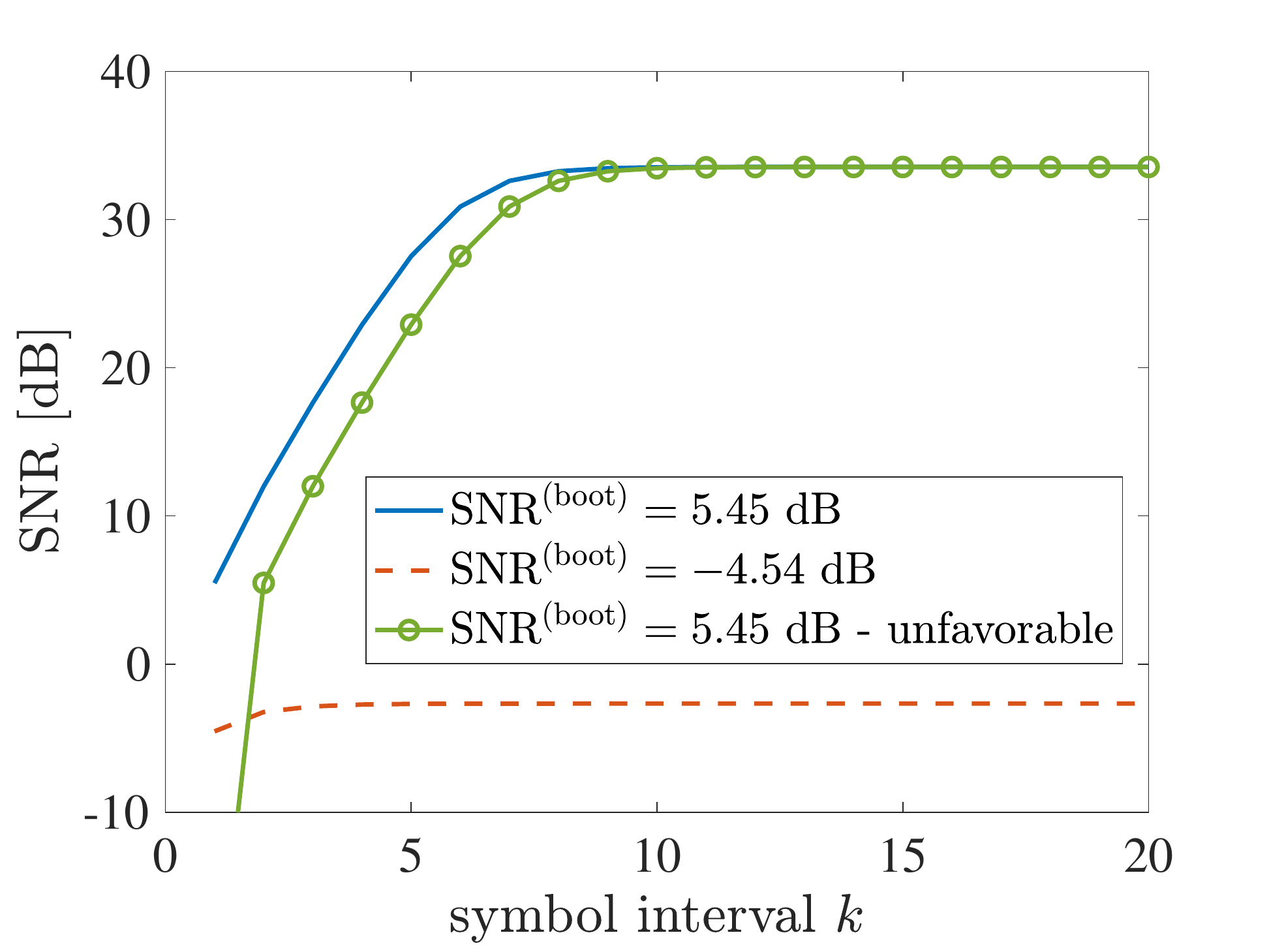}
		\caption{Time evolution of $\mathsf{SNR}_1[k]$ during the Scouting Task for different bootstrap conditions in the case of a rank-1 channel. }
		\label{Fig:SNRevol_v1}
\end{figure}

\begin{figure}[t!]
	\centering \includegraphics[trim= {0 0 0 0}, clip, width=0.8\linewidth]{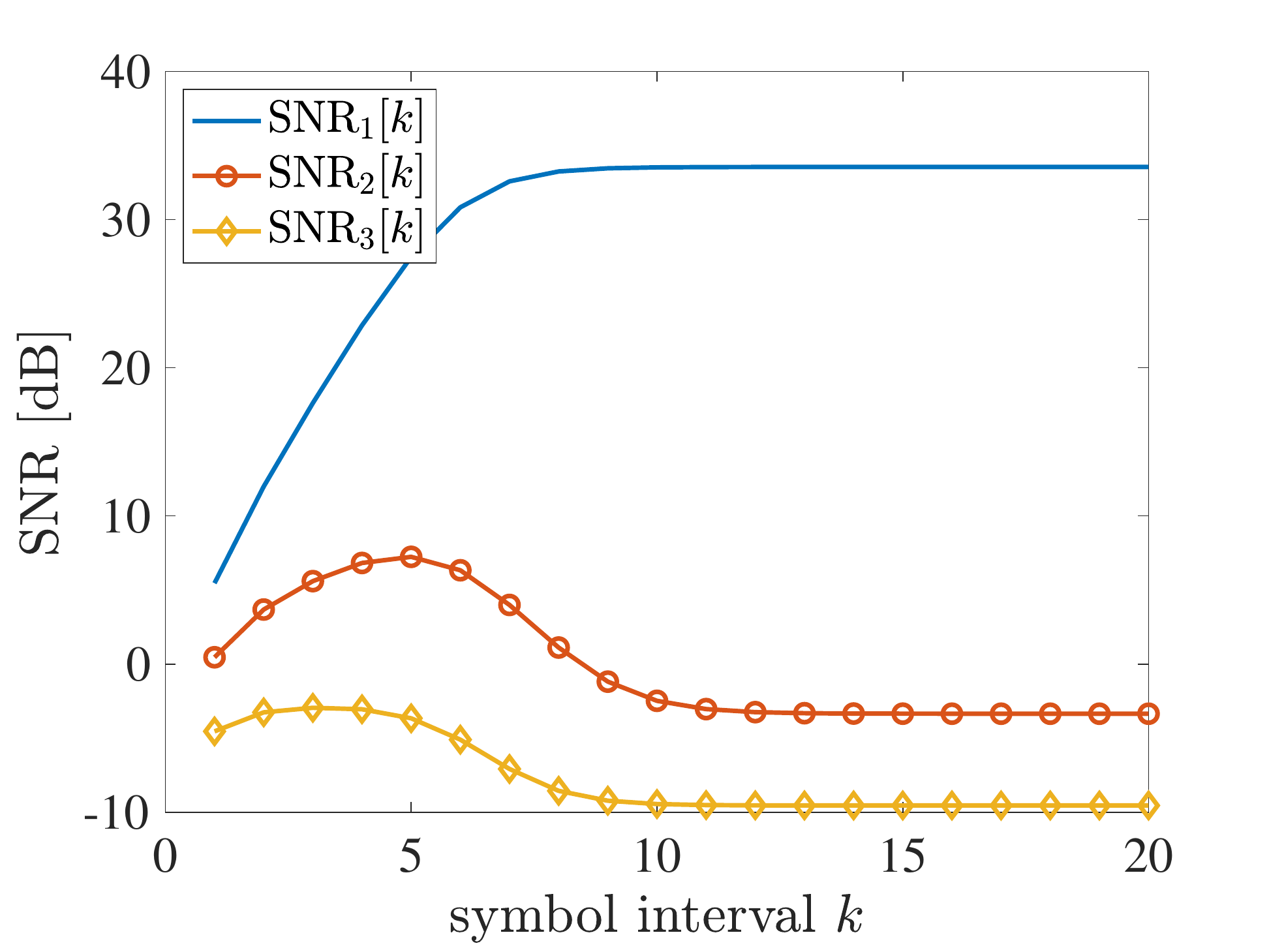}
		\caption{Time evolution of $\mathsf{SNR}_j[k]$ during the Scouting Task  in the case of a rank-3 channel. }
		\label{Fig:SNRevol_v3}
\end{figure}


\begin{figure}[t!]
	\centering \includegraphics[trim= {0 0 0 0}, clip, width=0.8\linewidth]{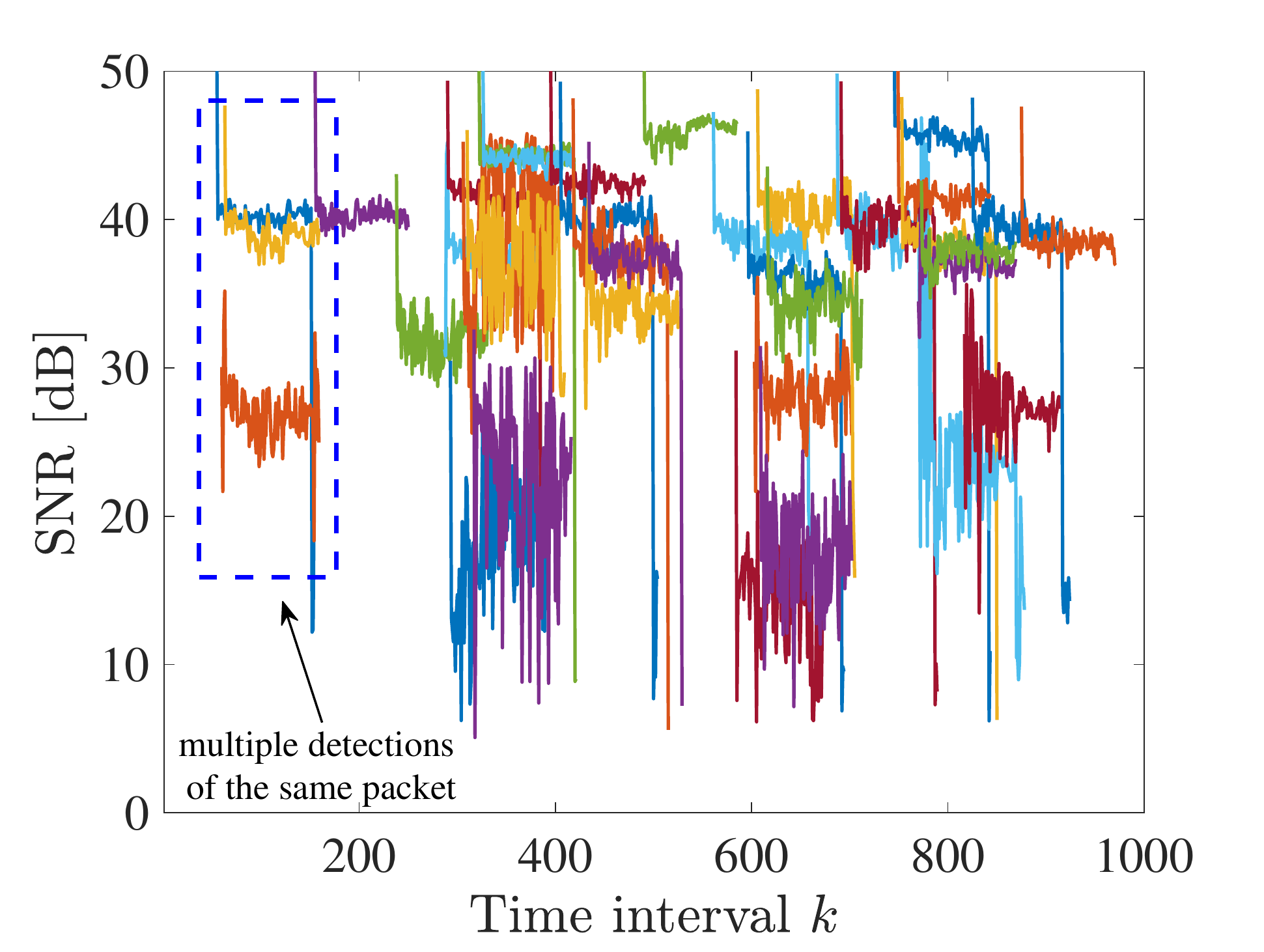}
		\caption{Example of SNR evolution during the Communication Tasks with $G=2$. Different colors refers to different packets/\acp{UE} managed by different Communication Tasks.}
		\label{Fig:snrG2}
\end{figure}

\begin{figure}[t!]
	\centering \includegraphics[trim= {0 0 0 0}, clip, width=0.8\linewidth]{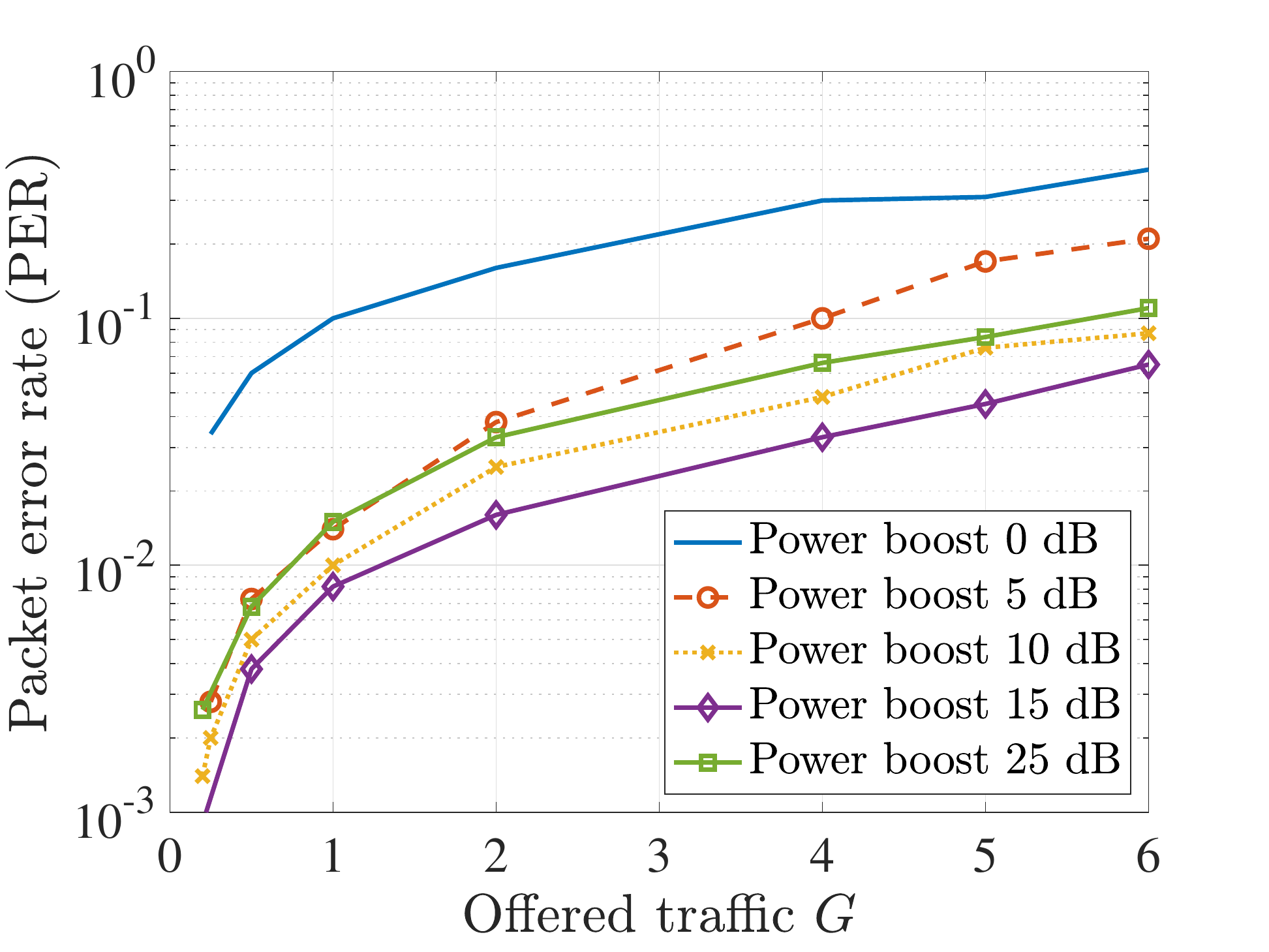}
		\caption{Packet error rate as a function of the offered traffic $G$. Effect of Scouting Task power boost. Free space condition.  $\Ptx=-5\,$dBm. }
		\label{Fig:PERboost}
\end{figure}

\begin{figure}[t!]
	\centering \includegraphics[trim= {0 0 0 0}, clip, width=0.8\linewidth]{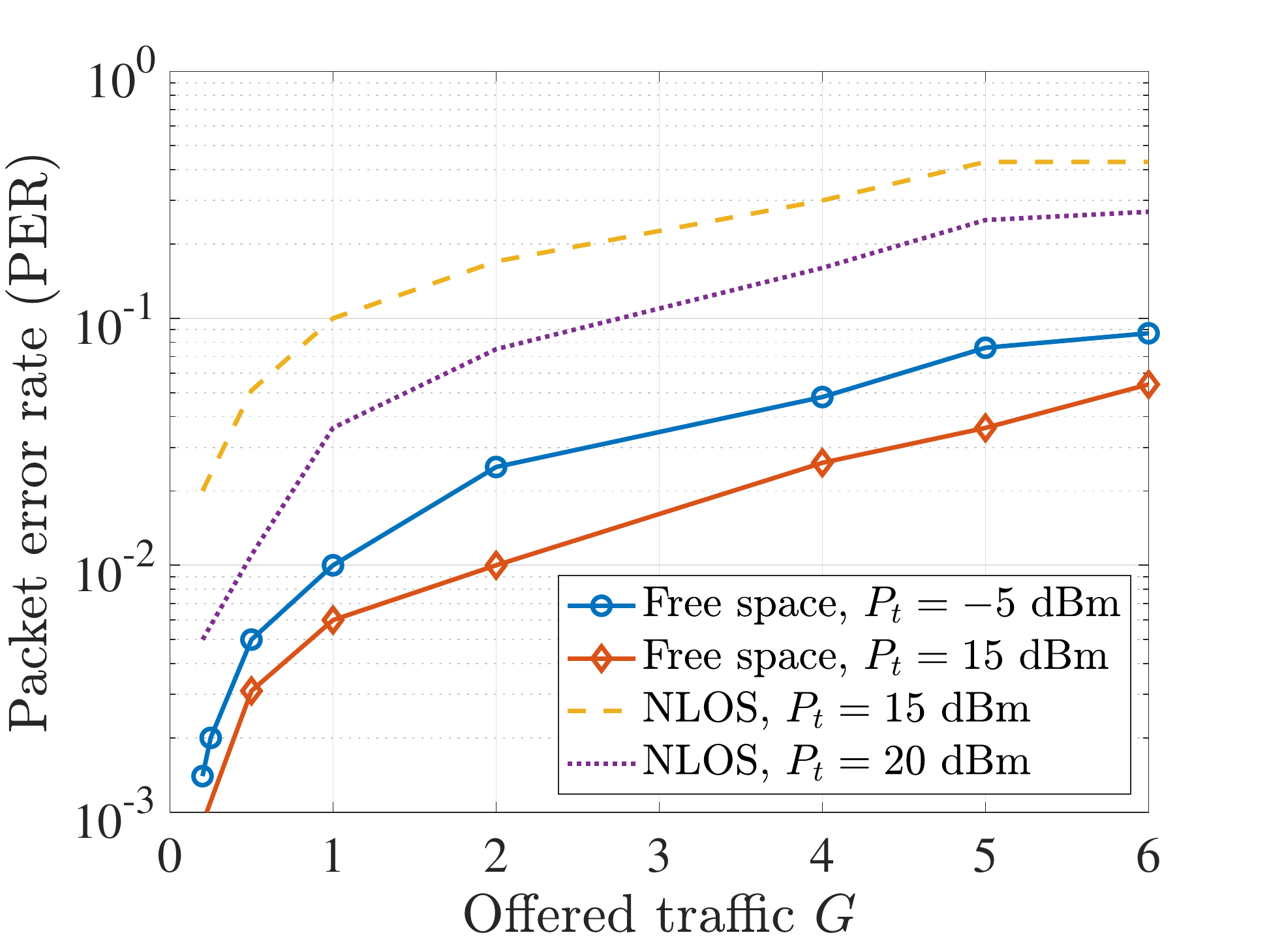}
		\caption{Packet error rate as a function of the offered traffic $G$ for different transmitted power levels and channel models. Power boost $10\,$dB.}
		\label{Fig:PER}
\end{figure}


\begin{figure}[t!]
	\centering \includegraphics[trim= {0 0 0 0}, clip, width=0.8\linewidth]{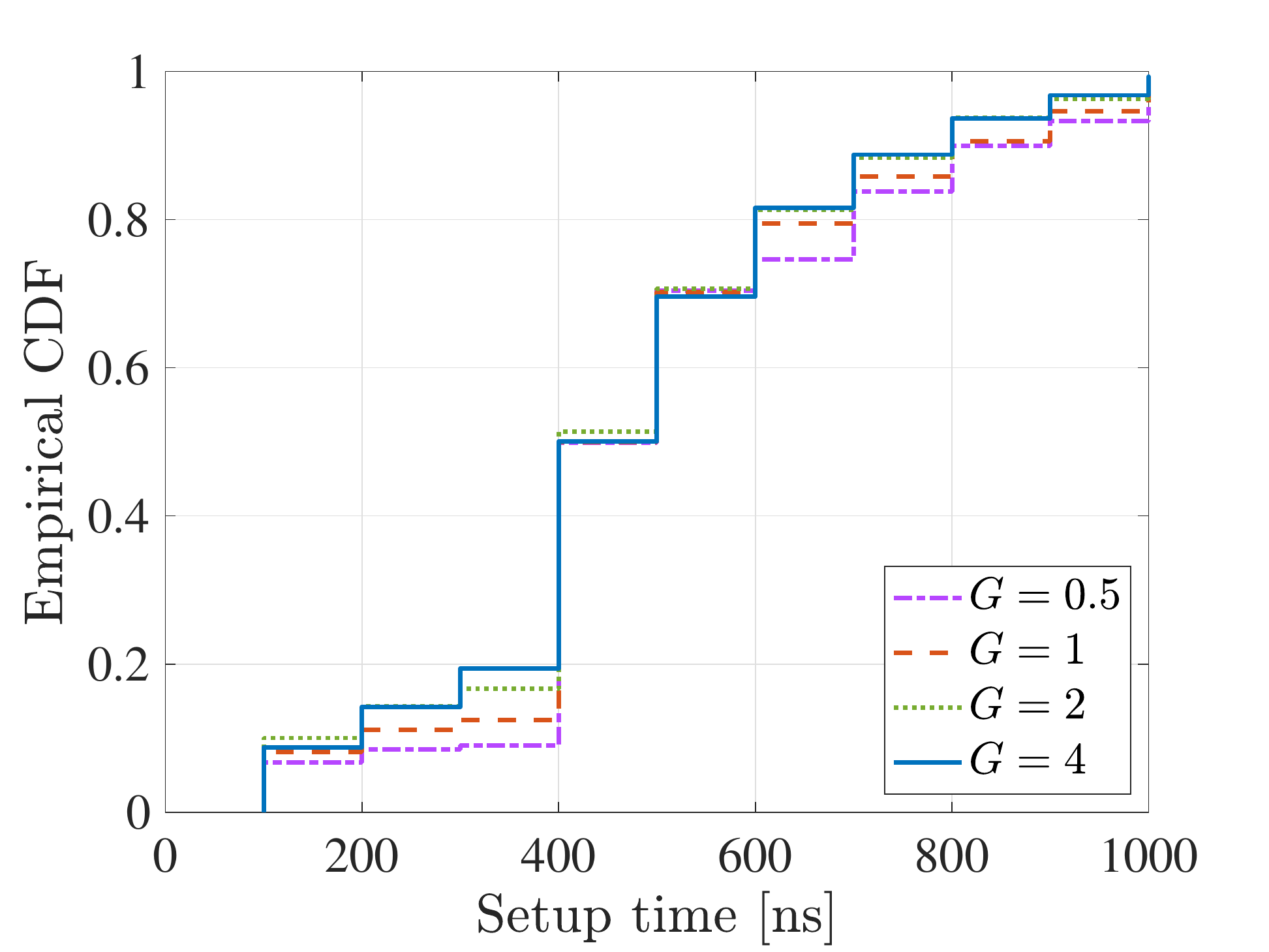}
		\caption{Empirical \ac{CDF} of the set up time for different values of the offered traffic. Free-space condition. }
		\label{Fig:cdftime}
\end{figure}

\section{Numerical Results}
\label{Sec:NumericalResults}



In this section, we report some simulation results with the purpose to investigate the performance  of the proposed scheme.
The values of the system parameters adopted for the analysis and the simulations are  reported in Table~\ref{Tab:Parameters}, if not otherwise specified. The scenario considered is composed of a \ac{BS}, equipped with a planar array deployed along the $xy$-plane,  located at $(0 ,0 ,0)\,$[m], 
and  \acp{UE} generating very short packets of total length $K=144\,$bits (32 bytes payload). In particular, the packets are generated randomly with exponentially distributed inter-arrival time with mean  $\mu\,$[s], corresponding to a total offered traffic $G=T_{\text{p}}/\mu$, where $T_{\text{p}}=K\, T$ is the packet duration. Each packet corresponds to an \ac{UE} whose position is generated randomly with uniform distribution within the area centered in $(0 ,0 ,10)\,$[m] and size $10 \times 10\,\left [\text{m}^2 \right ]$. The \ac{UE}'s \acp{SCM} are supposed to lay on the $xy$-plane.  

In Fig.~\ref{Fig:SNRevol_v1}, the time evolution of $\mathsf{SNR}_1[k]$ during the Scouting Task is shown for different bootstrap conditions in the case of rank-1 channel computed using the iterative equation \eqref{eq:SNRj1}, with $k^*=1$. In the first 2 plots, a random initial guess of the beamforming vector, corresponding to $\mathsf{SNR}_1[1] \simeq \mathsf{SNR}_1^{\text{(max)}}/N$, was assumed, whereas the third plot is related to an even more unfavorable situation where $\mathsf{SNR}_1[1]=\mathsf{SNR}_1^{\text{(max)}}/N^2$. 
The blue and green curves have been evaluated with $\mathsf{SNR}_1^{\text{(max)}}=35\,$dB and $\mathsf{SNR}_1^{\text{(max)}}=25\,$dB, respectively, to which a bootstrap \ac{SNR} of $5.45\,$dB and $-4.54\,$dB correspond. It can be noticed that when the bootstrap \ac{SNR} is larger than one, the \ac{SNR} converges after a few steps to $\mathsf{SNR}_1^{\text{(max)}}$ and the packet can be detected, whereas it converges to very low \ac{SNR} values when  the bootstrap \ac{SNR} is less than one, as predicted by \eqref{eq:sol}. From Fig.~\ref{Fig:SNRevol_v1}, it can also be concluded that the impact of the initial beamforming vector on the \ac{SNR} evolution is marginal and it does not affect the final value.

The situation with a rank-3 channel is analyzed in Fig.~\ref{Fig:SNRevol_v3} where the evolution of $\mathsf{SNR}_j[k]$, $j=1,2, 3$, is plotted for  $\mathsf{SNR}_1^{\text{(max)}}=35\,$dB, $30\,$dB, and $25\,$dB. The proposed iterative method ensures the convergence to the top eigenvector of the channel, as  can be appreciated in the figure (blue curve). 
Comparing the blue curves in Figs.~\ref{Fig:SNRevol_v1} and \ref{Fig:SNRevol_v3}, it can also be observed that the convergence speed to the top-eigenvector is only slightly affected by the rank of the channel. 
%
In all cases, the convergence time is below 10 time intervals. This suggests that 
a guard time of $K_{\text{g}}=16$ symbols set to zero,   inserted at the beginning of each packet and followed by the payload of 32 bytes, gives sufficient time to the Scouting Task to converge and detect the packet.

An example of realization of the \ac{SNR} evolution in a time frame of 1000 symbols is shown in Fig.~\ref{Fig:snrG2}  for an offered traffic of $G=2$ in free-space condition. In particular, the \ac{SNR} evolution during the Communication Task of each detected packet, identified by a different color, is plotted.  
The \ac{SNR} fluctuations are mainly due to the inter-user interference because the channels are not  orthogonal, i.e., the favorable propagation condition is not met. As it can be seen, despite most of the transmissions are overlapped, the scheme is capable of resolving them in an efficient manner. Different \acp{SNR} are due to  different path loss values (\acp{UE}' positions) and inter-user interference  caused by partial channels subspaces overlap, whereas the first peak is the residual of the power boost during the last scouting symbol. Highlighted with a blue rectangle are  multiple successful detections of the same packet by different Communication Tasks using different estimated eigenvectors. Such a situation is representative of  a near-field channel condition for the associated \ac{UE}.  

In Fig.~\ref{Fig:PERboost}, the \ac{PER}, corresponding to $1-\{\text{Probability of packet detection}\}$, as a function of the offered traffic $G$ is shown for different power boost levels in free space. 
The \ac{PER} has been computed through Monte Carlo simulations  terminated when at least 100 erroneous packets were counted. It can be noticed that the power boost  is in general beneficial but also that a compromise has to be found also avoiding too much higher values.  
In fact, while an increase of the power boost facilitates the detection of new packets by the Scouting Task, on the other hand, it might increase the interference to the packets already under decoding in the absence of favorable propagation. Simulations evidenced that $15\,$dB provides the lowest \ac{PER} in all cases, but the performance is still good with a lower level of $10\,$dB that is considered in the next plots.  
 Obviously, under favorable propagation (massive \ac{MIMO}) the inter-user interference is minimized and the determination of the power boost is less critical.

The impact of the transmitted power and channel models can be observed in Fig.~\ref{Fig:PER}. In particular, free-space and the \ac{NLOS} CDL-C multipath 3GPP channel models \cite{TR38.901:2019}   with a delay spread of $9\,$ns have been considered. Obviously, higher transmitted power levels  are necessary in \ac{NLOS} condition due to the more severe path loss.
The increase of the transmitted power  lowers slightly  the \ac{PER} indicating that the performance is mainly limited by the inter-user interference.
Achieving favorable propagation through a substantial increase of the antenna elements at the \ac{BS} is one possible solution to mitigate or even eliminate the  inter-user interference. Future investigations will be oriented to approaches based on successive interference cancellation schemes exploiting the packet diversity obtained in near-field conditions where the same packet is likely received associated with different eigenvectors. Another line of investigation is to combine the proposed scheme with coded random access schemes \cite{PaoSteLivPop:15} to further increase the reliability.   

Finally, Fig.~\ref{Fig:cdftime}  shows the empirical \ac{CDF} of the set-up time distribution of successfully detected and decoded packets, i.e., the convergence time of the Scouting Task,  for different values of the offered traffic. As it can be noticed, the impact of the offered traffic is marginal and the values are extremely low, below $1\, \mu$s, corresponding to 10 symbols.
It is worth noting that successfully decoded packets are subject to neither jitter nor latency as each  packet is available at the \ac{BS} as soon as the last bit has been transmitted and received, i.e., within the symbol time $T$. 

From the computational complexity point of view, the complexity of the proposed scheme is proportional to the number of parallel active Communication Tasks which, in turn, depends on the offered traffic. Both the Scouting and Communication Tasks involve  elementary operations whose number is proportional to the number $N$ of antennas at the \ac{BS} but not to the number of cells $M$ at the \ac{SCM}. Therefore, $M$ can be considered as a useful degree of freedom that can be exploited to improve the link budget, and hence the $\mathsf{SNR}^{(\text{boot})}$, without any drawback in terms of algorithm's complexity.


\section{Conclusion}
\label{Sec:Conclusions}

In this paper, we have proposed the adoption of modulating \acp{SCM} as a means to realize ultra-low complexity grant-free \ac{MIMO} communications exploiting retro-directive backscattering.
Thanks to the iterative algorithm  introduced in this paper, inspired by the Power Method, the \ac{BS} can estimate the optimal beamforming vector for the  \ac{BS}-\ac{UE} round-trip channel and hence establish the optimal single-layer \ac{MIMO} communication.
Numerical results have put in evidence that  \ac{MIMO} communications can be established  after a few iterations, with $\mu$s-level setup time and almost zero  latency and jitter,  even using very large arrays and in the presence of realistic channels characterized by multipath.
The proposed system is particularly appealing in many applications envisioned for 6G requiring low-complexity \ac{MIMO} solutions at high frequencies such as  \ac{IIoT}.

%


%


\ifCLASSOPTIONcaptionsoff
\fi
\bibliographystyle{IEEEtran}

\bibliography{Biblio/IEEEabrv,Biblio/MetaSurfaces,Biblio/IntelligentSurfaces,Biblio/MassiveMIMO,Biblio/MIMO,Biblio/THzComm,Biblio/Channels,Biblio/WINS-Books,Biblio/RandomAccessURLLC,Biblio/BiblioDD}

\end{document}